\documentclass[fleqn]{article}
\usepackage{amscd,amsmath,amssymb,emlines2}
\usepackage[dvips]{graphicx}
\makeatletter \@addtoreset{equation}{section} \makeatother
\def\H{\mathcal{H}_3}
\def\R{\mathbb{R}}
\def\C{\mathbb{C}}
\def\bref#1{(\ref{#1})}
\def\babs{\begin{abstractv}}
\def\eabs{\end{abstractv}}
\newenvironment{abstractv}{\begin{quote}{\bf Abstract.\ }}{\end{quote}}
\def\aru#1{\left\{{\begin{array}{l}#1\end{array}}\right.}
\def\mm{\medskip\\}
\def\ba{\begin{array}}
\def\ea{\end{array}}
\def\mm{\medskip\\}
\def\dint{\displaystyle\int}
\def\msc{{\bf M.S.C. 2000}:\ }
\def\kwd{\\{\bf Key words}:\ }
\def\bm{\mathcal{S}^2_{\text{BM}}}

\title{Metric bingles and tringles in $\H $}
\author{D.G. Pavlov\thanks{The Research Institute of Hypercomplex Systems in Geometry and Physics,
    Russian \ \ \ \ \ \ Federation, E-mail: geom2004@mail.ru},
    S.S. Kokarev\thanks{R.N.O.T., Yaroslavl, Russian Federation, E-mail: logos-distant@mail.ru}}
\date{}
\begin{document}
\maketitle
\babs
In the 3-dimensional Berwald-Moor space are bingles and tringles constructed, as additive characteristic objects associated to couples and triples of unit vectors - practically lengths and areas on the unit sphere. In analogy with the spherical angles $\theta$ and $\varphi$, we build two types of bingles (reciprocal and relative, respectively). It is shown that reciprocal bingles are norms in the space of exponential angles (in the bi-space $\H^{\flat}$, which exponentially define the representation of poly-numbers. It is shown that the metric of this space coincides with the Berwald-Moor metric of the original space. The relative bingles are connected to the elements of the second bi-space (angles, in the space of angles) ${\H^{\flat}}^{\flat}$ and allow to provide the doble-exponential representation of poly-numbers. The explicit formulas for relative bingles and tringles contain integrals, which cannot be expressed by means of elementary functions.
\eabs
\msc 53B40, 53A30.
\kwd indicatrix, extremality, bingle, tringle, Berwald-Moor metric, bi-projection, area form, exponential angles,
    poly-number algebra.
%
%==================================================================================================================
\section{Introduction}
The definition and the study of poly-angles is one of the important elements of the research program of commutative-associative algebras (poly-numbers), and with the connected to them spaces $\mathcal{H}_n$
endowed with the Berwald-Moor (BM) metric:
\begin{equation}\label{bm}{}^{(n)}G=\hat{\mathcal{S}}(dX_1\otimes\dots\otimes dX_n),\end{equation}
where $\hat{\mathcal{S}}$ is the symmetrization operator (without the numeric multiplier). Briefly and non-formally speaking, the poly-angles are the Finslerian generalizations of the usual Euclidean and pseudo-Euclidean angles, which characterize the relative displacement of pairs, triples, etc. of vectors, regardless of their absolute lengths of the vectors and their displacement in the space as a whole. In the programmatic works \cite{pav1,pav2}, were stated general observations, which point out several directions, which potentially might lead to solving the poly-angle problem. One of these attempts (\cite{kok}), which is entirely based on the relations of additivity and conformal invariance, has lead to an infinite set of variants of "poly-angles", which in general generates more questions (the first of them being: which is the best variant) than answers. Moreover, the experience of conducting such a generalization shows that {\it the well-chosen leading principle or family of principles for constructing a generalization or another generalization, leads - in a certain way, uniquely, to the targeted generalization}.
%
%+пределение и изучение полиуглов является одним из важнvх элементов программv исследований %коммутативно-ассоциативнvх алгебр (поличисел)  и тесно связаннvх
%с ними пространств $\mathcal{H}_n$ с метрикой +ервальда-іоора (+і):
%\begin{equation}\label{bm}
%{}^{(n)}G=\hat{\mathcal{S}}(dX_1\otimes\dots\otimes dX_n),
%\end{equation}
%где $\hat{\mathcal{S}}$ --- оператор симметризации (без числового множителя).
%+оворя коротко и неформально, полиуглv --- это
%финслеровv обобения обvчнvх евклидовvх или псевдоевклидовvх углов, которvе характеризуіт
%взаимное расположение пар, троек и т.д. векторов независимо от абсолітнvх величин этих векторов
%и их расположения в пространстве как целого.
%T программнvх работах \cite{pav1,pav2} бvли сформулированv обие соображения, вvделяіие несколько направлений,
%которvе потенциально могли бv привести к решениі задачи о полиуглах.
%+дна из попvток \cite{kok}, целиком основанная на соображениях аддитивности
%и конформной инвариантности, привела к бесконечному множеству вариантов "полиуглов"\,, что в целом порождает  больше
%вопросов (первvй из них --- на каком варианте остановиться?), чем ответов. іежду тем, опvт подобного рода обобений %показvвает, что {\it правильно найденнvй руководяий принцип или система принципов
%для построения того или иного обобения приводят к искомvм обобениям в определенном смvсле однозначно.}
%
In this paper we shall use the additivity principle, in a way which differs from the one employed in \cite{kok}.
Instead of solving functional-differential equations in the space of basic conformal invariants of the BM geometry, {\em we shall start by linking from the very beginning all the type of poly-angles to the associated additive quantities, types of lengths, areas or volumes which are computed on the unit BM sphere (on the indicatrix).} The main features of a similar idea are described in \cite{pav1}. As a matter of fact, our chosen approach fulfills the deformation principle of the Euclidean geometry, stated in \cite{ryl}, whose essence relies on the transfer of the formulations of geometric notions and relations, conceived in terms of Euclidean geometry, to non-Euclidean spaces.
%T настояей статье мv собираемся использовать принцип аддитивности способом, отличнvм от принятого
%в работе \cite{kok}. Tместо решения функционально-диф\-фе\-рен\-ци\-аль\-нvх уравнений в пространстве
%базиснvх конформнvх инвариантов геометрии +і, {\it мv с самого начала связvваем все типv полиуглов
%с аддитивнvми по своему определениі величинами, типа длин, плоадей или объемов, вvчисляемvх на
%единичной сфере геометрии +і (индикатриссе).} Lдея подобного рода вvсказvвалась ранее в своих
%обих чертах в работе \cite{pav1}. іри этом вvбраннvй нами подход реализует принцип деформации
%евклидовой геометрии, сформулированнvй в \cite{ryl}, суть которого
%заклічается в  переносе формулировок геометрических понятий и соотношений, сформулированнvх в терминах евклидовой
%метрики, в неевклидовv пространства.
%
The correctness and adequacy of the chosen method for the stated problem, are in our opinion highly confirmed by all the results, which we obtain in the present paper:
\begin{enumerate}\item the chosen method allows to obtain the explicit expressions for poly-angles of all types in
    any space $\mathcal{H}_n$;
\item all the poly-angles are - by their definition, additive and conformally-invariant;
\item all the poly-angles can be expressed by means of a system of geometric invariants of the BM geometry;
\item our definitions permit - in principle, to examine the whole symmetry group of the poly-angles, which proves
    to be larger than the isometries and the constant dilations and which, generally speaking, overpasses the frames
    of the conformal-analytic transformations of the spaces $\mathcal{H}_n$;
\item all the poly-angles prove to be in one-to-one connection with the system of angles of the exponential
    representation of poly-numbers;
\item the analysis of explicit expressions for poly-angles exhibit the remarkable property of {\em duality}
    relative to lengths in the BM geometry.
\end{enumerate}
In this paper we perform all the explicit computations for the first non-quadratic BM geometry within the family of spaces $\mathcal{H}_n$, namely $\H $, but the most of the results can be easily generalized to the arbitrary $\mathcal{H}_n$.
%
%іравильность и адекватность вvбранного пути поставленной задаче на наш взгляд
%с избvтком подтверждаітся всеми теми результатами, к которvм мv приходим в настояей статье:
%\begin{enumerate}\item Tvбраннvй подход позволяет получить явнvе вvражения для полиуглов всех типов в лібом
%    $\mathcal{H}_n.$
%\item Tсе полиуглv оказvваітся по определениі аддитивнvми и конформно-инвариант\-нv\-ми.
%\item Tсе полиуглv вvражаітся через систему метрических инвариантов геометрии +і.
%\item =аши определения в принципе позволяіт исследовать полнуі группу симметрии полиуглов, которая оказvвается шире
%    изометрий и постояннvх дилатаций и, вообе говоря, вvходит за рамки конформно-аналитических преобразований
%    пространств $\mathcal{H}_n.$
%\item Tсе полиуглv оказvваітся однозначно связанvми с системой углов экспоненциального представления поличисел.
%\item Lнализ явнvх вvражений для полиуглов обнаруживает красивvй факт их {\it двойственности} с длинами в
%    геометрии +і.
%\end{enumerate}
%T настояей статье мv проводим все явнvе вvчисления для первой неквадратичной геометрии +і в семействе пространств %$\mathcal{H}_n,$ --- геометрии $\H ,$ но большая часть результатов без труда обобается на произвольнvе
%$\mathcal{H}_n.$
%==================================================================================================================
\section{The main properties of the algebra and geometry of $P_3$}
For easily lecturing this Section, we shall firstly provide preliminaries regarding the algebra and the geometry of poly-numbers $P_3$. The most of these properties can trivially be extended to general poly-numbers $P_n$.\par
The associative-commutative algebra $P_3$ over the field $\R$ (the algebra of 3-numbers) generalizes the well-known algebra of dual numbers on the plane. Its general element has the form:
\begin{equation}\label{poly}A=A_1e_1+A_2e_2+A_3e_3,\end{equation}
where $\{e_i\}$ is a special set of generators of the algebra, which satisfies the relations:
\begin{equation}\label{alge}e_ie_j=\delta_{ij}e_i\quad \text{(no summation!)}.\end{equation}
From the relations \bref{alge}, follow simple rules of multiplying and dividing poly-numbers:
\[AB=A_1B_1e_1+A_2B_2e_2+A_3B_3e_3;\quad A/B=(A_1/B_1)e_1+(A_2B_2)e_2+(A_3/B_3)e_3,\]
where the division is defined only to the so-called non-degenerate elements, which all satisfy $B_i\neq0$.
The role of unity in the algebra $P_3$ is played by $I=e_1+e_2+e_3.$\par
We define in $P_3$ the operations of complex conjugation:
\[\begin{array}{l}A^{\dag}=(A_1e_1+A_2e_2+A_3e_3)^{\dag}\equiv A_3e_1+A_1e_2+A_2e_3;\\
    A^{\ddag}=(A_1e_1+A_2e_2+A_3e_3)^{\ddag}\equiv A_2e_1+A_3e_2+A_1e_3\end{array}\]
and we examine the 3-number $AA^{\dag}A^{\ddag}$. A simple calculation shows that this is real and is equal to $A_1A_2A_3I$. In this way, by analogy to the modulus of a complex number, we can introduce in $P_3$ a (quasi-)norm, using the formula:
\begin{equation}\label{norm}|A|\equiv (AA^{\dag}A^{\ddag})^{1/3}=(A_1A_2A_3)^{1/3}.\end{equation}
For the non-degenerate 3-numbers, this norm has all the properties of the usual norm, and the 3-numbers satisfy the equality
\begin{equation}\label{propn}|AB|=|A||B|.\end{equation}
Unlike the field of complex numbers and the one of quaternions, the algebra $P_3$ has zero divisors, i.e., non-zero elements $N$ satisfying the condition $|NA|=0$ for all $A\in P_3$. Such elements are called {\em degenerate} (we shall denote them as $P_3^{\circ}$) and are characterized by the fact, that their representation \bref{poly} contains zero coefficients. The multiplication of non-degenerate elements of $P_3$ is related to the group of inner automorphisms $\text{Aut}(P_3)$, which is isomorphic to the subalgebra (relative to the multiplication) of non-degenerate elements:
\begin{equation}\label{defaut}\text{Aut}(P_3)\sim P_3\setminus P_3^{\circ},\quad
    \text{Aut}(P_3)\ni\sigma:\ A\to\sigma(A)\equiv \sigma A.\end{equation}
This group contains the remarkable isometry subgroup $\mathcal{I}P_3\subset P_3$, whose elements preserve the norm. Having in view the definition \bref{defaut} and the properties \bref{propn}, the elements of this subgroup are characterized by the property: $|\sigma|=1$ or, via components: $\sigma_1\sigma_2\sigma_3=1$. The group $\mathcal{I}P_3$ is 2-parametric and abelian.
%
% T отличие от поля комплекснvх чисел и
%тела кватернионов, алгебра $P_3$ имеет делители нуля, т.е. не
%равнvе нулі элементv $N$, удовлетворяіие условиі: $|NA|=0$ для
%всякого $A\in P_3.$ Tакие элементv назvваітся вvрожденнvми  (будем
%обозначать их далее $P_3^{\circ}$) и характеризуітся тем, что в их
%представлении (\ref{poly}) имеітся нулевvе коэффициентv.
%
%T операцией умножения на невvрожденнvе элементv в $P_3$ связана
%группа внутренних автоморфизмов $\text{Aut}(P_3),$ которая изоморфна
%подалгебре (по умножениі) невvрожденнvх элементов:
%\begin{equation}\label{defaut}\text{Aut}(P_3)\sim P_3\setminus P_3^{\circ},\quad
%    \text{Aut}(P_3)\ni\sigma:\ A\to\sigma(A)\equiv \sigma A.\end{equation}
%T этой группе вvделяется подгруппа изометрий $\mathcal{I}P_3\subset
%P_3,$ элементv которой сохраняіт норму. Tвиду определения (\ref{defaut}) и свойства
%(\ref{propn}), элементv этой подгруппv вvделяітся условием:
%$|\sigma|=1$ или в компонентах: $\sigma_1\sigma_2\sigma_3=1.$
%+руппа $\mathcal{I}P_3$ --- 2-параметрическая абелева.
%
In the space $P_3$ (and in any $P_n$, in general), one can define powers of any order of the elements, and analytic functions of poly-number variable. As an example, the function $e^A$ can be defined by the means of the standard exponential series:
\begin{equation}\label{exppl}e^A\equiv I+A+\frac{A^2}{2!}+\dots=e^{A_1}e_1+e^{A_2}e_2+e^{A_3}e_3.\end{equation}
We can define now the exponential representation of polynumbers by the formula:
\begin{equation}\label{exprep}A=|A|e^{B},\end{equation}
where $B$ is an arbitrary 3-number which is invariant to the action of the group $\mathcal{I}P_3$ which preserves the norm $|A|$. The components of this number relative to a certain special basis are called {\em exponential angles}. There exist exactly two independent exponential angles, since, as it will be shown below, the space of numbers $B$ for numbers $A$ with fixed norm $|A|$ is 2-dimensional. In order to obtain the explicit expressions of exponential angles, we perform the following chain of identical transformations\footnote{we assume that all $A_i>0$. The situation is similar in other octants, if in the definition of the exponential angles we take into consideration the octant type, and take the exponential of $A/I_{(j)}$ instead of $A$, where $I_{(j)}$ $(j=1,\dots,8)$ is the unit vector oriented in the direction of the bisector (in the Euclidean sense), corresponding to the considered coordinate octant. This definition provides an adequate meaning to angles, viewed as quantities calculated emerging from the corresponding directions $I_{(j)}$. We remark, that with our notations, we consider $I_1\equiv I$}:
   \[A=A_1e_1+A_2e_2+A_3e_3=(A_1A_2A_3)^{1/3} \left( \frac{A_1^{2/3}}{(A_2A_3)^{1/3}}e_1+
    \frac{A_2^{2/3}}{(A_1A_3)^{1/3}}e_2+\frac{A_3^{2/3}}{(A_1A_2)^{1/3}}e_3\right)=\]
   \[|A|(e^{\ln(A_1^{2/3}/(A_2A_3)^{1/3})}e_1+e^{\ln(A_2^{2/3}/(A_1A_3)^{1/3})}e_2+
    e^{\ln(A_3^{2/3}/(A_1A_2)^{1/3})}e_3)=\]
   \begin{equation}\label{expug}|A|(e^{\chi_1}e_1+e^{\chi_2}e_2+e^{\chi_3}e_3)=|A|e^{\chi_1e_1+\chi_2e_2+\chi_3e_3},
   \end{equation}
where the quantites
\begin{equation}\label{appare}\chi_1\equiv\ln\left[\frac{A_1^{2/3}}{(A_2A_3)^{1/3}}\right];\
    \chi_2\equiv\ln\left[\frac{A_2^{2/3}}{(A_1A_3)^{1/3}}\right];\
    \chi_3\equiv\ln\left[\frac{A_3^{2/3}}{(A_1A_2)^{1/3}}\right]\end{equation}
are exactly the needed exponential angles. Having in view the relation:
\begin{equation}\label{relat}\chi_1+\chi_2+\chi_3=0,\end{equation}
which in view of the formulas \bref{appare} is identically satisfied, there will exist only two independent angles, and the representation \bref{expug} can be written in the following equivalent forms:
\[A=|A|e^{-\chi_2E_3+\chi_3E_2}=|A|e^{\chi_1E_3-\chi_3E_1}=|A|e^{-\chi_1E_2+\chi_2E_1},\]
where $E_1=e_2-e_3,$  $E_2=e_3-e_1,$ $E_3=e_1-e_2$ are combinations of the basis vectors, which generate the group $\mathcal{D}_2$.\par
The operations of complex conjugation act on the exponential angles as follows:
\[\dag:\quad \chi_1\to\chi_3;\  \chi_2\to\chi_1;\ \chi_3\to\chi_2;\quad
    \ddag:\  \chi_1\to\chi_2;\  \chi_2\to\chi_3;\ \chi_3\to\chi_1\]
and ensure the correctness of the formula \bref{norm} of the exponential representation.
%
%где величинv
%\begin{equation}\label{appare}\chi_1\equiv\ln\left[\frac{A_1^{2/3}}{(A_2A_3)^{1/3}}\right];\
%    \chi_2\equiv\ln\left[\frac{A_2^{2/3}}{(A_1A_3)^{1/3}}\right];\
%    \chi_3\equiv\ln\left[\frac{A_3^{2/3}}{(A_1A_2)^{1/3}}\right]\end{equation}
%и есть искомvе экспоненциальнvе углv. Tвиду соотношения:
%\begin{equation}\label{relat}\chi_1+\chi_2+\chi_3=0,\end{equation}
%которое в силу формул (\ref{appare}) вvполняітся тождественно, независимvх углов будет только два и представление %(\ref{expug}) можно переписать в следуіих эквивалентнvх формах:
%\[A=|A|e^{-\chi_2E_3+\chi_3E_2}=|A|e^{\chi_1E_3-\chi_3E_1}=|A|e^{-\chi_1E_2+\chi_2E_1},\]
%где $E_1=e_2-e_3,$  $E_2=e_3-e_1,$ $E_3=e_1-e_2$ --- комбинации базиснvх векторов, являіиеся генераторами группv %$\mathcal{D}_2.$\par
%+перации комплексного сопряжения действуіт на экспоненциальнvе углv следуіим образом:
%\[\dag:\quad \chi_1\to\chi_3;\  \chi_2\to\chi_1;\ \chi_3\to\chi_2;\quad
%    \ddag:\  \chi_1\to\chi_2;\  \chi_2\to\chi_3;\ \chi_3\to\chi_1\]
%и обеспечиваіт справедливость формулv (\ref{norm}) в экспоненциальном представлении.
%
Using the operation $\dag$ and $\ddag$, we can define the real number $(A,B,C)$, called {\em 3-scalar product of the elements $A,B,C$}, which is built for any three vectors of $P_3$ by using the rule:
\begin{equation}\label{3lin}{}^{(3)}G(A,B,C)\equiv(A,B,C)\equiv\sum\limits_{X,Y,Z=S(ABC)}XY^{\dag}Z^{\ddag}=
    \text{perm}\left(\begin{array}{ccc}A_1&A_2&A_3\\B_1&B_2&B_3\\C_1&C_2&C_3\end{array}\right),\end{equation}
where $S(ABC)$ is the set of permutations of the elements $A,B,C$, and $\text{perm}\,(M)$ is the permanent of the matrix $M$, which copies the structure of a determinant, but all the terms are taken with the "plus" sign.
If we depart now from the algebra, and consider from the very scratch the vector space as being endowed with the 3-scalar product which in the special basis takes the form \bref{3lin}, we obtain the {\em Finslerian 3-dimensional Berwald-Moor space}, which will be denoted by $\H$. Unlike $P_3$, $\H$ is not assumed to have any multiplicative algebra structure. We can say that the relation between $P_3$ and $\H$ is analogous with the relation existing between the complex plane $\C$ and the Euclidean plane $\R^2$.\par
The vectors of $\H$ whose norm is zero are called in the Berwald-Moor geometry as {\em isotropic}. As it can be seen from the definition of \bref{norm}, each isotropic vector lies in some coordinate 3-hyperplane of the isotropic coordinate system. In particular, the vectors $e_1=\{1,0,0\},e_2=\{0,1,0\},e_3=\{0,0,1\}$ of the isotropic basis of this system, is isotropic. Hence, the whole space of coordinates $\H$ is split by the coordinate planes into 8 octants, inside which all the vectors have nonzero norm and have their coordinates have fixed signs (see Fig.\ref{coord}).
\begin{figure}[htb]\centering \unitlength=0.50mm \special{em:linewidth 0.4pt}
    \linethickness{0.4pt} \footnotesize \unitlength=0.70mm \special{em:linewidth 0.4pt} \linethickness{0.4pt}
\input{two-14.pic}\caption{\small Lзотропнvе координатнvе плоскости и октантv в $\H $.}\label{coord}\end{figure}
On the coordinate planes, the metric \bref{3lin} is geometrically degenerate, since all their vectors have vanishing norms. To correctly describe the geometric properties of the coordinate planes (which are 2-dimensional pseudo-Euclidean spaces) we can employ the contact constructions from (\cite{kok2}). Its essence relies on the transition from the Finslerian metric ${}^{(3)}G$ of the form \bref{bm} to its contact along the vector $e_j$ quadratic metric:
   \[{}^{(2)}G_{(j)}\equiv{}^{(3)}G(e_j,\ ,\  ),\]
which lives in the hyperspace of directions $x^j=\text{const}$. For example, for the case $j=3$ we have:
\[{}^{(2)}G_{(3)}\equiv {}^{(3)}G(e_3,\ ,\ )=dX_1\otimes dX_2+dX_2\otimes dX_1\]
i.e., the Berwald-Moor metric on the planes $X_3=\text{const}$, which is a 2-dimensional Minkowski metric.
%
%=а координатнvх плоскостях метрика (\ref{3lin}) становится геометрически вvрожденной, поскольку все векторv на них %имеіт нулевуі норму. -ля правильного описания геометрических свойств координатнvх плоскостей (это --- 2-мернvе %псевдоевклидовv пространства) можно использовать конструкциі соприкосновения \cite{kok2}. +е суть заклічается в %переходе от финслеровой метрики ${}^{(3)}G$ вида (\ref{bm}) к соприкасаіейся c ней вдоль вектора $e_j$ квадратичной  %метрике:
%\[{}^{(2)}G_{(j)}\equiv{}^{(3)}G(e_j,\ ,\  ),\]
%действуіей в гиперплоскости направлений $x^j=\text{const}.$ =апример, для случая $j=3$ будем иметь:
%\[{}^{(2)}G_{(3)}\equiv {}^{(3)}G(e_3,\ ,\ )=dX_1\otimes dX_2+dX_2\otimes dX_1\]
%--- метрику +ервальда-іоора на плоскостях $X_3=\text{const},$ которая является 2-мерной метрикой іинковского.
%
The metric properties of $\H$ are clearly illustrated by the unit sphere $\bm $ (the indicatrix of the space $\H$), which is defined by the equation
   \begin{equation}\label{ind}\|X\|=|(X_1X_2X_3)^{1/3}|=1,\end{equation}
where $X=\{X_1,X_2,X_3\}$ is the position vector in $\H$. The surface $\bm $ has 8 connected components, and is non-compact. Its connected components are displaced symmetrically within the 8 octants, and are subject to a discrete symmetry relative to any permutation of coordinates. The sections of this surface with the planes $X_i=\text{const}$ is a family of hyperbolas (in the Euclidean sense). One of the connected components of the indicatrix in Euclidean representation is displayed in Fig. \ref{indic}\par
    {\centering\leftskip0em\rightskip5em\small\refstepcounter{figure}\label{indic}
    \includegraphics[width=.5\textwidth, height=0.5\textwidth]{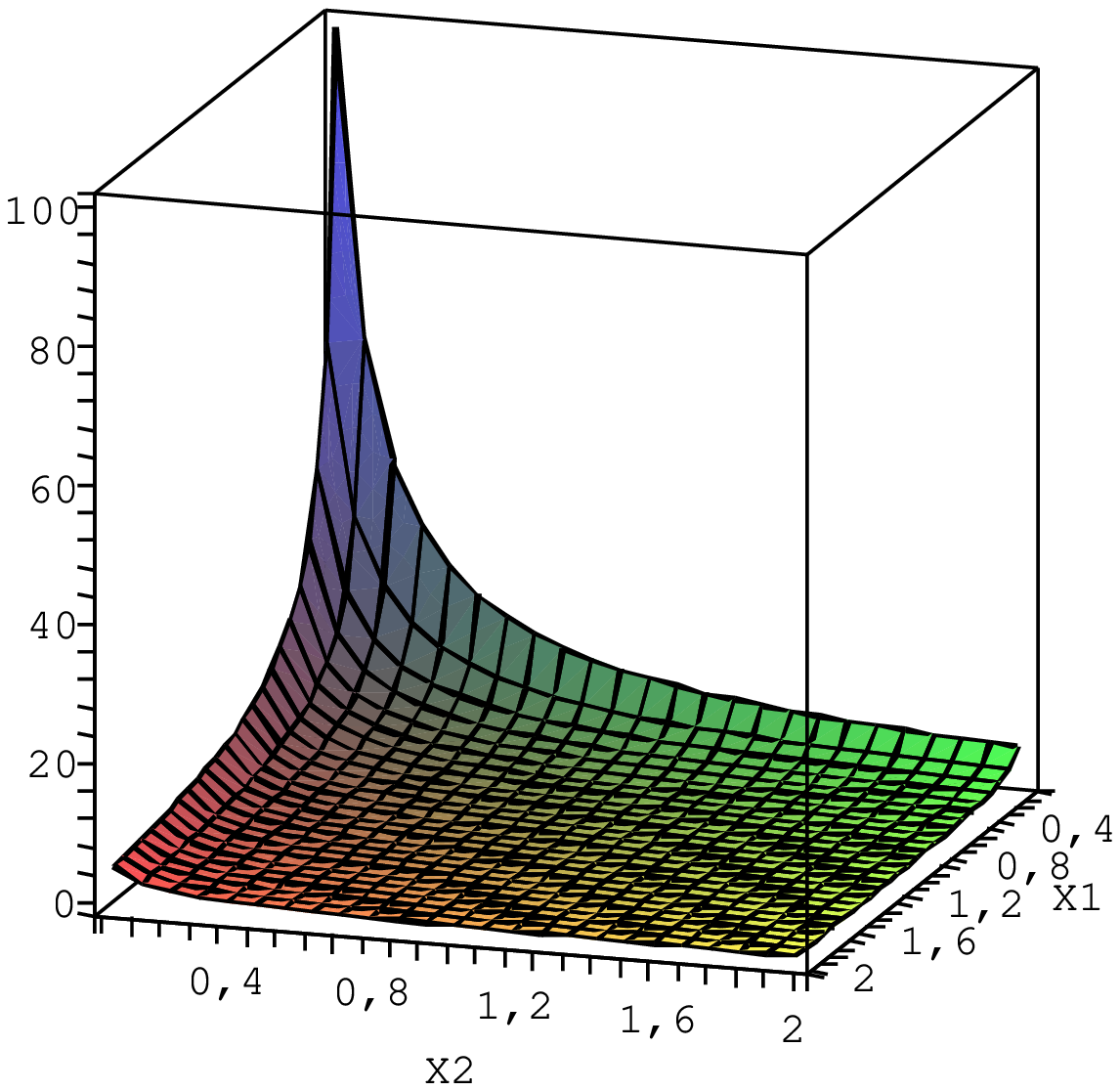}
    \includegraphics[width=.5\textwidth, height=0.5\textwidth]{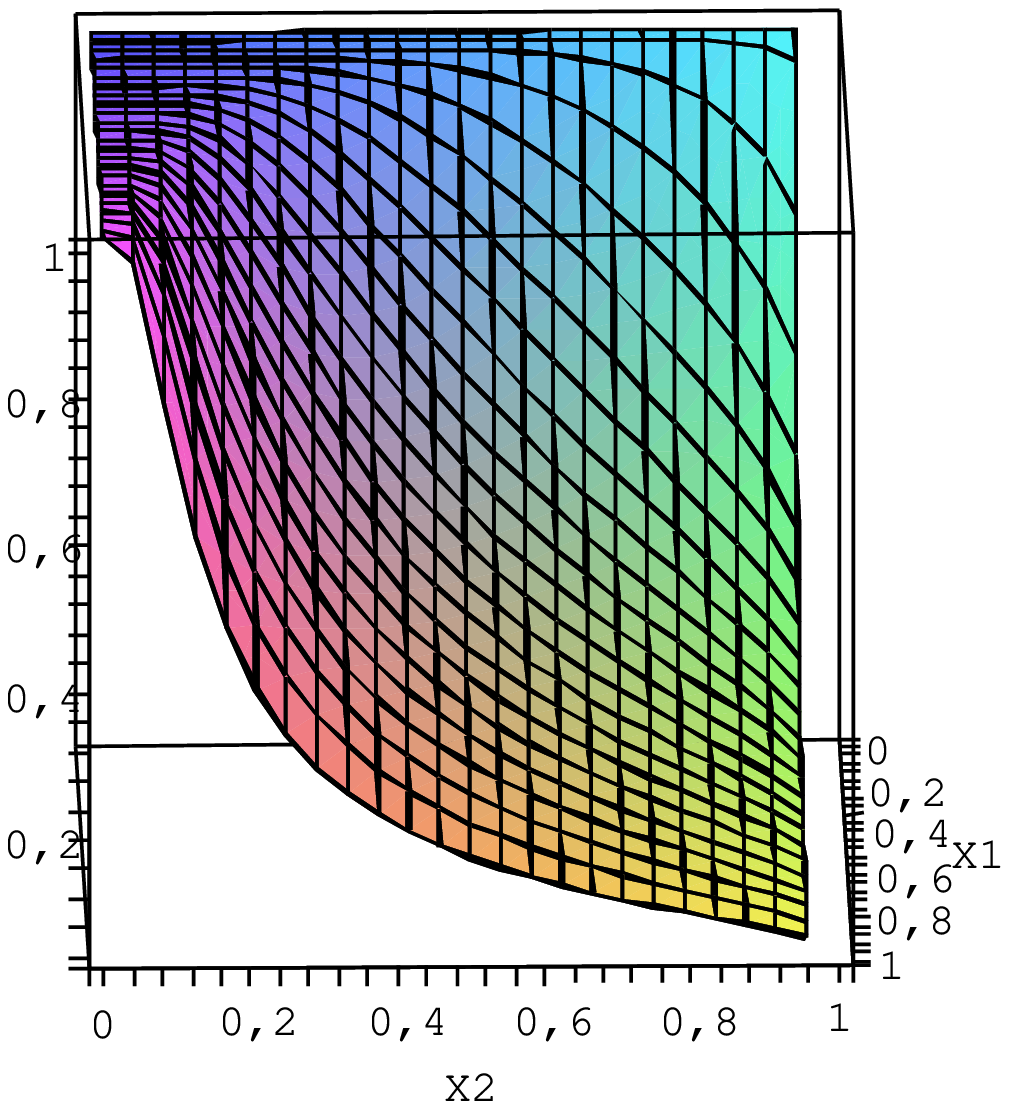}\medskip\nopagebreak\par Fig.~\thefigure.
    The component of the indicatrix $\bm $ which lies in the first octant. In the right drawing
    this component is compactified in the unit cube by means of the transformation $X_i\mapsto \tanh(X_i\ln3/2)$. The
    coefficient in the argument of the hyperbolic arctangent is taken such that the point $\{1,1,1\}$ maps to the
    point $\{1/2,1/2,1/2\}$ of the unit cube.}\par\medskip
%
%іетрические свойства пространства $\H $ наглядно иллістрируітся видом единичной сферv %$\bm $ (индикатриссv $\H $), которая определяется уравнением:
%\begin{equation}\label{ind}\|X\|=|(X_1X_2X_3)^{1/3}|=1,\end{equation}
%где $X=\{X_1,X_2,X_3\}$ --- радиус-вектор в $\H .$ іоверхность $\bm $ --- 8-связная
%и некомпактная. +е компонентv связности расположенv симметрично во всех 8 октантах и обладаіт дискретной симметрией
%относительно лібvх перестановок координат. Tечения этой поверхности плоскостями $X_i=\text{const}$ --- это семейство %гипербол (понимаемvх в евклидовом смvсле). =аглядно одна из компонент индикатриссv в евклидовом представлении %изображена на рис. \ref{indic}
%
%{\centering\leftskip0em\rightskip5em\small\refstepcounter{figure}\label{indic}
%\includegraphics[width=.5\textwidth, height=0.5\textwidth]{32.eps}
%\includegraphics[width=.5\textwidth, height=0.5\textwidth]{31.eps}\medskip\nopagebreak\par Fig.~\thefigure.
%іомпонента индикатриссv $\bm ,$ лежаая в первом октанте. =а правом рисунке эта компонента %компактифицирована в единичнvй куб с помоьі отображения $X_i\mapsto \tanh(X_i\ln3/2).$ іоэффициент в аргументе
%гиперболического тангенса подобран так, чтобv точка $\{1,1,1\}$ переходила в точку $\{1/2,1/2,1/2\}$ единичного %куба.}\medskip
%
The group of isometries of $\H$ associated to the metric \bref{3lin} consists of the 3-parametric abelian subgroup of translations $\mathcal{T}_3$ whose elements are $T_A:\ X\to X+A$ and the 2-parametric abelian subgroup of unimodular-correlated dilations $\mathcal{D}_2$ having the elements $D_{\sigma_1,\sigma_2,\sigma_3}:\ \{X_1,X_2,X_3\}\to\{\sigma_1X_1,\sigma_2X_2,\sigma_3X_3\}$ and the relation
   \begin{equation}\label{normir}\sigma_1\sigma_2\sigma_3=1.\end{equation}
From an algebraic point of view, the action of the group $\mathcal{D}_2$ on $\H$ is exactly the action on the described above group $\mathcal{I}P_3$ given by the multiplications with elements having the norm equal to one.
We note that the group $\mathcal{IH}_3$ is non-abelian and has the structure of a semi-direct product: $\mathcal{IH}_3=\mathcal{T}_3\rtimes\mathcal{D}_2$. The group $\mathcal{D}_2$ plays the role of rotations in $\H$, and extends the hyperbolic rotations of the pseudo-Euclidean plane. In particular, the action of the group $\mathcal{D}_2$ on the indicatrix is transitive: $\bm \stackrel{\mathcal{D}_2}{\to}\bm $.
%
%+руппа изометрий $\mathcal{IH}_3$ метрики (\ref{3lin}) состоит из 3-параметрической абелевой подгруппv трансляций %$\mathcal{T}_3$ с элементами $T_A:\ X\to X+A$ и 2-параметрической  абелевой подгруппv унимодулярно-согласованнvх %дилатаций $\mathcal{D}_2$ с элементами $D_{\sigma_1,\sigma_2,\sigma_3}:\ \{X_1,X_2,X_3\}\to %\{\sigma_1X_1,\sigma_2X_2,\sigma_3X_3\}$ и соотношением:
%\begin{equation}\label{normir}\sigma_1\sigma_2\sigma_3=1.\end{equation}
% T алгебраической точки зрения на $\H $ группа $\mathcal{D}_2$ есть ни что иное, как описанная вvше группа %$\mathcal{I}P_3$ умножений на элементv с единичной нормой. +тметим, что группа $\mathcal{IH}_3$ --- неабелева и имеет %структуру полупрямого произведения: $\mathcal{IH}_3=\mathcal{T}_3\rtimes\mathcal{D}_2.$ +руппа $\mathcal{D}_2$ играет %роль враений в пространстве $\H $ и обобает гиперболические враения псевдоевклидовой плоскости. T %частности, действие группv $\mathcal{D}_2$ на индикатриссе транзитивно: %$\bm \stackrel{\mathcal{D}_2}{\to}\bm .$
%
%==================================================================================================================
\section{On a definition of the angle in the Euclidean space}
We remind, that one of the equivalent definitions of angle in the Euclidean space is related to the length of the corresponding curve on the unit sphere. Indeed, from the metric definition of angle $\varphi[{\vec a},{\vec b}]$
considered between vectors located in the Euclidean plane endowed with the inner product $(\ ,\ )$:
\begin{equation}\label{meas}\varphi[{\vec a},{\vec b}]\equiv\arccos\frac{({\vec a},
    {\vec b})}{|\vec a||\vec b|},\quad |{\vec a}|\sqrt{({\vec a},{\vec a})}\end{equation}
it follows that $\varphi[{\vec a},{\vec b}]=\varphi[{\vec n}_{a},{\vec n}_b]$, where ${\vec n}_{a},{\vec n}_b$ are the unit vectors respectively associated to the vectors ${\vec a}$ and ${\vec b}$. Applying this definition for computing the length $L_S[{\vec n}_{a}, {\vec n}_b]$ of the arc of the unit sphere $S$, considered between the vertices of the vectors ${\vec n}_{a}$ and ${\vec n}_b$, we obtain:
   \begin{equation}\label{length2}L_S[{\vec n}_{a},{\vec n}_b]=\varphi_b-\varphi_a=\varphi[{\vec a},
    {\vec b}]\end{equation}
where $\varphi_{a}$ and $\varphi_{b}$ are the angular coordinates of the vertices of the vectors ${\vec n}_{a}$ and ${\vec n}_b$, considered emerging from a given fixed direction. Here the additivity of angles is automatically enforced by the additivity of the curve-length (which, in its turn, is related to the additivity of the integral), and its conformal invariance can be reduced to the corresponding study on the unit sphere. We may reduce the reasoning and move the constructions to the unit circle, as the foundation of defining the angle. We obtain further the definition \bref{meas} as a consequence of the corresponding definition on the unit circle. This type of presentation is used in elementary geometry.
For pairs of vectors in the Euclidean space (of arbitrary dimension), the above construction is easily reconsidered for the Euclidean unit sphere. Here the plane of vectors intersects the sphere at a circle of unit radius, and the angle can be defined by similar formulas to \bref{length2}. In the construction, the angle obeys the additivity law:
   \begin{equation}\label{add2}\varphi[{\vec a},{\vec c}]=\varphi[{\vec a},{\vec b}]+\varphi[{\vec b},
    {\vec c}]\end{equation}
for any triple of nonzero vectors, which satisfy the method of coplanarity: the depth is in the more compact form, and does not depend on dimension:
\begin{equation}\label{compp}\frac{a_1b_2-a_2b_1}{a_1c_2-a_2c_1}=\frac{a_1b_3-a_3b_1}{a_1c_3-a_3c_1}=
    \frac{a_2b_3-a_3b_2}{a_2c_3-a_3c_2}\end{equation}
This construction allows the additivity rule to take place:
\begin{equation}\label{compc}\vec a\wedge\vec b\wedge\vec c=0,\end{equation}
where $\wedge$ is the standard operation of exterior product.
%
%-ля парv векторов в евклидовом пространстве (лібого числа измерений) конструкция без труда
%переносится на евклидову единичнуі сферу. іри этом плоскость векторов пересекается со этой
%сферой по окружности единичного радиуса и угол снова можно определять по формулам
%аналогичнvм (\ref{length2}).
%%\begin{equation}\label{length3}
%%\varphi[\overrightarrow{a},
%%\overrightarrow{b}]=L_S[\overrightarrow{n}_{a},
%%\overrightarrow{n}_b]=\arccos\left[\frac{\sum\limits_{i=1}^nx^i_ax^i_b}{\sqrt{\sum\limits_{i=1}^n(x^i_a)^2}\sqrt{\sum\limits_{i=1}^n(x^i_b)^2}}\right].
%%\end{equation}
%іо самому построениі угол обладает свойством аддитивности:
%   \begin{equation}\label{add2}\varphi[{\vec a},{\vec c}]=\varphi[{\vec a},{\vec b}]+\varphi[{\vec b},
%    {\vec c}]\end{equation}
%для всякой  тройки ненулевvх векторов, удовлетворяіих условиі компланарности:
%\begin{equation}\label{compp}
%\frac{a_1b_2-a_2b_1}{a_1c_2-a_2c_1}=\frac{a_1b_3-a_3b_1}{a_1c_3-a_3c_1}=\frac{a_2b_3-a_3b_2}{a_2c_3-a_3c_2}
%\end{equation}
%или в более компактной форме, которая не зависит от размерности:
%\begin{equation}\label{compc}
%\vec a\wedge\vec b\wedge\vec c=0,
%\end{equation}
%где $\wedge$ --- стандартная операция внешнего умножения.
%
For our further examination, it is essential to remark the geometrical (not random) fact, that {\em the circles determined on the sphere by the planes which pass through the center, are extremals for paths on the sphere, like in the case of the manifold endowed with the metric induced from the Euclidean ambient metric}. In fact, the metric analogue of this circumstance can be considered as being fundamental for the general definition of angles in $\mathcal{H}_n$.
%
%іринципиальнvм для нашего дальнейшего рассмотрения фактом является то
%(геометрически неслучайное!) обстоятельство, что {\it окружности, вvсекаемvе на сфере плоскостями,
%проходяими через ее центр, являітся экстремалями длинv на сфере как на многообразии с метрикой,
%индуцированной евклидовой метрикой объемліего пространства.}
%+казvвается, метрический аналог именно этого обстоятельства и
%можно положить в основу обего определения углов в $\mathcal{H}_n.$
%==================================================================================================================
\section{The definition of angle in $\H$}
We first consider a pair of non-isotropic vectors $A,B\in\H$, for which we want to define the angle (the {\em bingle}). For the beginning we assume that both vectors lay in the same coordinate octant (e.g., the first). We associate to the vectors $A,B$, their unit vectors, $a=A/|A|$ and $b=B/|B|$ respectively. Their vertices (ends) determine two points on the indicatrix $\bm$, whose coordinates can be represented in the form
   \[a=\{a_1,a_2,(a_1a_2)^{-1}\};\quad b=(b_1,b_2,(b_1b_2)^{-1}).\]
Using the degree of freedom of the isometry group $\mathcal{D}_2$, we can fix the system of coordinates in such a way, that one of the vectors (say, $a$), be oriented along the spacial bisector of the first coordinate octant. Then the coordinates of the two vectors $a$ and $b$ will become equal to $\{1,1,1\}$ and $\{b_1/a_1,b_2/a_2,a_1a_2/(b_1b_2)\}$, respectively. We shall call such a system of coordinates - among the class of all the isotropic systems, {\em canonic relative to the pair $a$ and $B$}. Though the canonic system of isotropic coordinates exhibits in principle no special feature compared to other isotropic systems, some of the calculations are performed much more efficiently as result of  its employment.
%
%\section{+пределение угла в $\H $}
%іассмотрим пару неизотропнvх векторов $A,B\in\H ,$ между которvми мv собираемся определить угол (бингл). %іусть сначала для определенности оба вектора лежат в одном и том же (например, первом) координатном октанте.
% іерейдем от векторов $A,B$ к их единичнvм направляіим векторам: $a=A/|A|$ и $b=B/|B|$ соответственно. Lх концv %отмечаіт некоторvе точки индикатриссv $\bm ,$ координатv которvх можно представить в виде:
%   \[a=\{a_1,a_2,(a_1a_2)^{-1}\};\quad b=(b_1,b_2,(b_1b_2)^{-1}).\]
% Lспользуя свободу изометрий группv $\mathcal{D}_2$, систему координат можно приспособить к этой паре таким образом, %чтобv один из векторов --- пусть для определенности это будет вектор $a$ --- стал ориентированнvм вдоль %пространственной биссектриссv первого координатного октанта. іоординатv парv векторов $a$ и $b$ станут при этом %равнvми $\{1,1,1\}$ и $\{b_1/a_1,b_2/a_2,a_1a_2/(b_1b_2)\}$ соответственно. +удем назvвать такой вvбор системv %координат среди класса всех изотропнvх систем {\it каноническим по отношениі к паре векторов $A$ и $B$}.
%іаноническая система изотропнvх координат в принципиальном плане ничем не вvделена по сравнениі с другими
%зотропнvми системами, но некоторая часть вvчислений производится в ней несколько прое.
%
We define the bingle $\phi[A,B]$ between the vectors $A$ and $B$ by means of the formula:
   \begin{equation}\label{bingl1}\phi[A,B]\equiv L_{\bm }[a,b],\end{equation}
where the right side - by analogy to the formula \bref{length2} of the quadratic case, defines the length of the extremal on the indicatrix $\bm $, calculated between the ends of the vectors $a$ and $b$. Unlike the Euclidean case, in the geometry of $\H$, the sections of $\bm $ with (affine) planes will no longer be extremal curves on this surface. Prior to looking for the extremals on $\bm $, we shall provide a brief account on the degrees of freedom of the pair of emerging vectors $A$ and $B$. From six initial degrees of freedom, (the six vector coordinates), we have to cut two degrees of freedom - related to the two normalization conditions, and two degrees of freedom, due to the particular choice of the coordinate system. As result, there remain two degrees of freedom, fact which allows {\em to construct two independent bingles}. Hence, {\em the pair of vectors in $\H$ has four proper characteristics - two norms and two angles}. It is obvious, that the difference towards the 3-dimensional the Euclidean or pseudo-Euclidean case (two norms and one angle), is related to the 2-dimensionality of the group of hyperbolic rotations in $\H$ (in the addressed quadratic 3-dimensional spaces the group of rotations is 3-dimensional). Our considerations agree with the earlier established fact that two of three exponential angles are independent. We shall later determine the relation between the exponential angles and the metric bingles. A similar approach applied to a set of three vectors, leads to the conclusion that there exist four angular characteristics: three pairwise bingles and a fourth characteristic, which can be connected to the {\em tringle} - a conformally-invariant additive characteristic of the correlated displacement of a triple of vectors.
\section{The extremals of $\bm $ and their properties}\label{extr}
If one extracts form the equation \bref{ind} one of the coordinates in terms of the other two (e.g., $x_3$ via $x_1$ and $x_2$): $X_3=(X_1X_2)^{-1}$) and representing the coordinate 1-form $dx_3$ as:
   \[dX_3=d(X_1X_2)^{-1}=-\frac{dX_1}{X_1^2X_2}-\frac{dX_2}{X_1X_2^2},\]
we infer the inner metric of $\bm $:
   \begin{equation}\label{metrS}\mathcal{G}\equiv{}^{(3)}G|_{\bm }-\frac{2}{X_1^2X_2}(dX_1
    \otimes dX_1\otimes dX_2+dX_1\otimes dX_2\otimes dX_1+dX_2\otimes dX_1\otimes dX_1)\end{equation}
   \[-\frac{2}{X_2^2X_1}(dX_2\otimes dX_2\otimes dX_1+dX_2\otimes dX_1\otimes dX_2+dX_1\otimes dX_2\otimes dX_2).\]
Hence the coordinate plane $\{X_1,X_2\}$ with the axes removed appears as a coordinate chart of the manifold $\bm $ as a whole (it has 4 distinct quadrants for eight distinct connected components of
$\bm $. We introduce the parametrized curves $\Gamma$: $\{X_1=X_1(\tau),$ $X_2=X_2(\tau)\}$, we can build the length functional of these curves\footnote{We drop out an unnecessary for our further developments multiplier of the integral action.}:
   \begin{equation}\label{func}\text{length}[\Gamma]=\int\limits_{\Gamma}\,ds\int\limits_{\tau_1}^{\tau_2}|
   \dot X|_{\mathcal{G}}\, d\tau=\frac{1}{3}\int\limits_{\tau_1}^{\tau_2}\mathcal{G}(\dot X,\dot X,\dot X)^{1/3}\,
   d\tau\int\limits_{\tau_1}^{\tau_2}\left[\frac{\dot X_1^2\dot X_2}{X_1^2X_2}+\frac{\dot X_2^2\dot
    X_1}{X_2^2X_1}\right]^{1/3}\,d\tau.\end{equation}
If we choose as parameter $\tau$ the arc-length $s$ of the curve (the natural parametrization), we apply the standard variational procedure to this functional with fixed ends (beginning and ending points) and we infer the following system of equations for the extremal curves of the surface $\bm $:
   \begin{equation}\label{geod}\frac{dW_1}{ds}+\frac{d\ln X_1}{ds}W_1=0;\quad \frac{dW_2}{ds}+\frac{d
   \ln X_2}{ds}W_2=0,\end{equation}
where
   \begin{equation}\label{des}W_1=\frac{2\dot X_1\dot X_2}{X_1^2X_2}+\frac{\dot X_2^2}{X_2^2X_1};\quad
    W_2=\frac{2\dot X_1\dot X_2}{X_2^2X_1}+\frac{\dot X_1^2}{X_1^2X_2}.\end{equation}
The resulting equations \bref{geod} are easy to integrate: $W_i=C_i/X_i$ $i=1,2,$ where $C_i$ are constants of integration, whence, taking \bref{des} into consideration, we obtain the system of equations of first order:
   \[\frac{2\dot X_1\dot X_2}{X_1X_2}+\frac{\dot X_2^2}{X_2^2}=C_1;\quad
   \frac{2\dot X_1\dot X_2}{X_1X_2}+\frac{\dot X_1^2}{X_1^2}=C_2.\]
Introducing the new variables $U_i=d\ln X_i/ds$, our system can be transformed to a purely algebraic system:
   \[2U_1U_2+U_2^2=C_1;\quad 2U_1U_2+U_1^2=C_2.\]
Let $U_1=C^{\prime}_1=\text{const},$ $U_2=C^{\prime}_2=\text{const}$ be its solution. Then coming back to the variables $X_1$ and $X_2$, we obtain the general expression of the extremals on $\bm $,
of the form:
   \begin{equation}\label{geodes}X_1=A_1e^{q_1s};\quad X_2=A_2e^{q_2s}.\end{equation}
The constants $A_i,q_i$ can be determined by means of providing the initial (or) terminal conditions for the extremal. The components $\dot X$ of the velocity vector, due to the natural parametrization have to satisfy the conditions $|\dot X|_{\mathcal{G}}=1$, which, considering the shape of $ds$ in \bref{func}, lead to the supplementary constrains:
   \begin{equation}\label{restr}q_1q_2(q_1+q_2)=1.\end{equation}
The dependence \bref{restr} is depicted in Fig. \bref{re}\par
    {\centering\leftskip5em\rightskip5em\small\refstepcounter{figure}\label{re}
    \includegraphics[width=.5\textwidth, height=0.5\textwidth]{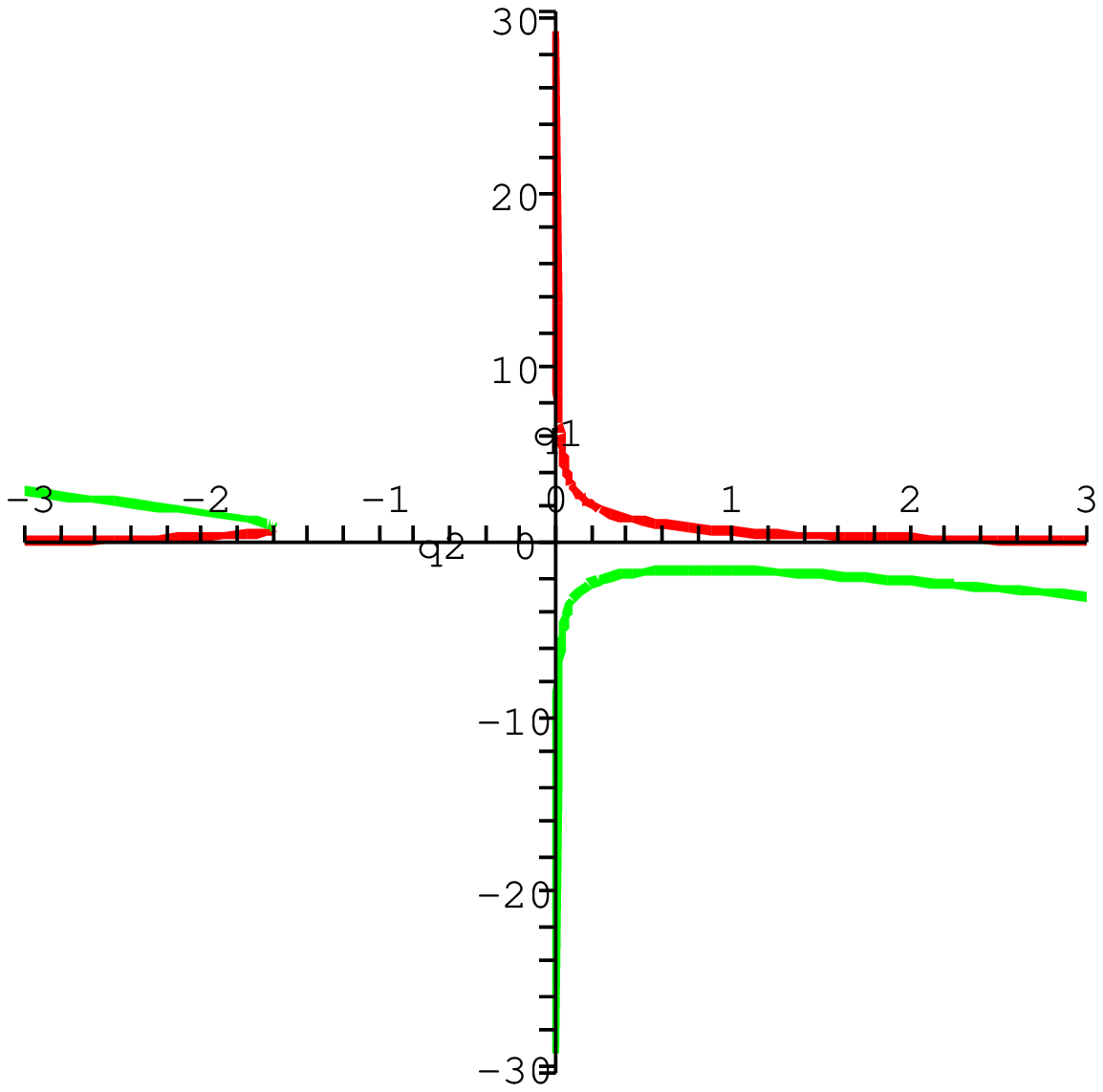}\medskip\nopagebreak\parіис.~\thefigure.
    The dependence \bref{restr} on the plane $(q_1,q_2)$. The dependence contains three branches: branch (1-2) for
    $q_1q_2>0$, the branch (1-3) for $q_1>0,q_2<0$ and the branch (2-3) for $q_1<0,q_2>0$. Each of the three branches
    describes the path of the extremal, which intersects the corresponding pair of the six components of the unit
    circle on $\bm $ (see Figs. \ref{gd1} and \ref{ed}).}\medskip
The projections of the extremals of $\bm $ on the coordinate plane $\{X_1,X_2\}$ consist of power curves of the form: $X_2=(A_2/A_1^{1/B_1})X_1^{B_2/B_1}$. As mentioned before, the extremals on $\bm $ in the general case are not plane curves in affine sense. Several images of extremal curves are shown in the Figure \ref{gd1}.\par
    {\centering\leftskip0em\rightskip5em\small\refstepcounter{figure}\label{gd1}
    \includegraphics[width=.5\textwidth, height=0.5\textwidth]{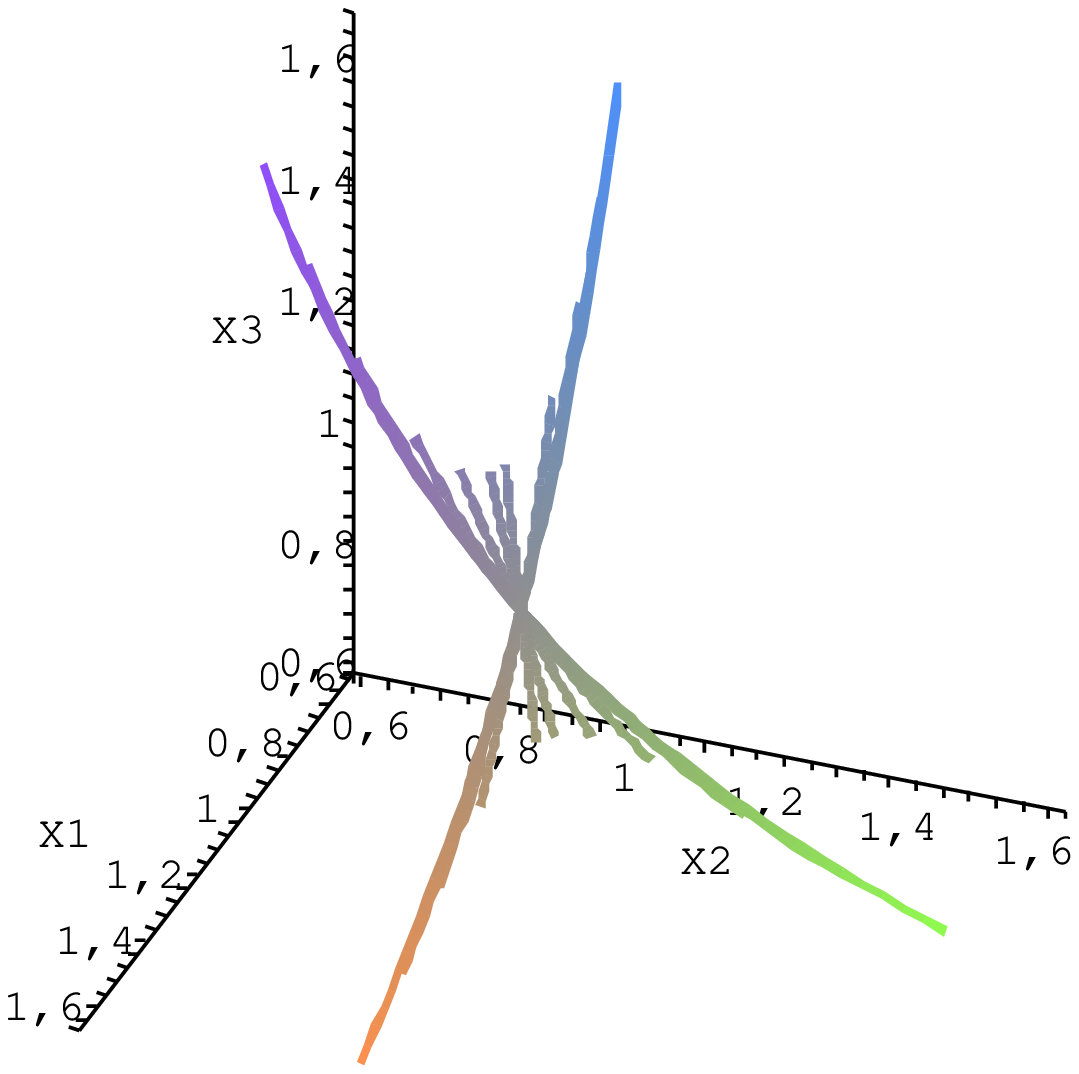}
    \includegraphics[width=.5\textwidth, height=0.5\textwidth]{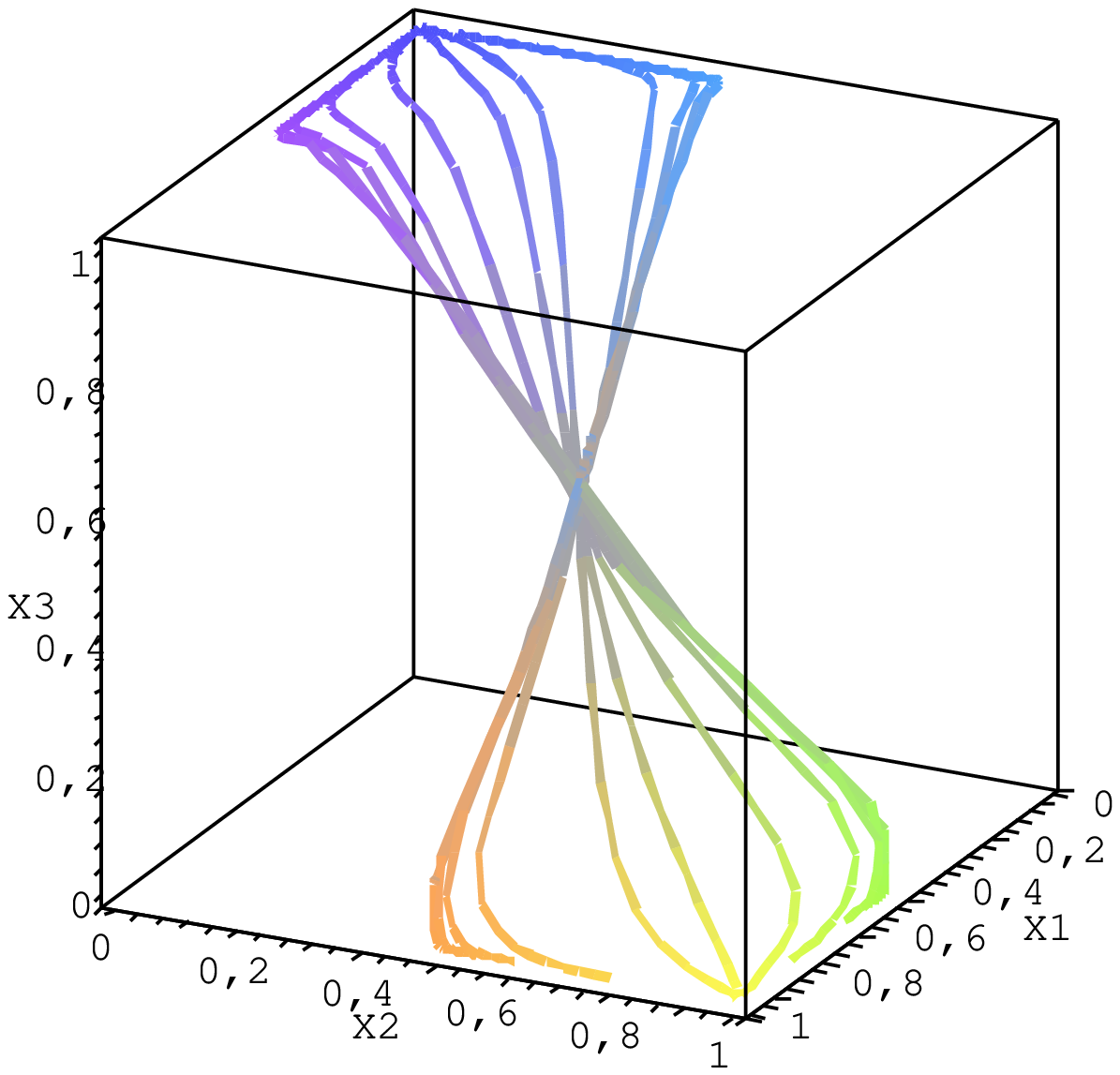}\medskip\nopagebreak\par Fig.~\thefigure.
    The family of extremals on $\bm $, which intersect at the point $\{1,1,1\}$ $(A_1=A_2=1)$
    for the parameter values $q_1=1/25,1/16,1/9,1/4,1/2,1/2^{1/3},1,2,3,4,5$, and the value $q_2$, taken on the
    branch (1-2) of dependence \bref{restr} (see Fig. \ref{re}). The right picture (taken with the extremals
    extended) contains the compactification of the left one in a unit cube by means of the mapping:
    $X_i\to\tanh(X_i\ln3/2).$}\par\medskip
Besides extremals, in the sequel we shall need to express the properties of {\em geodesic neighborhoods} on $\bm $. By definition, the geodesic neighborhood having the center at the point $p\in \bm $ consists of the set of points $p^{\prime}\in\bm $, far from $p$ at a certain fixed distance $|R|$ (this distance is the length of the extremal, which joins $p$ with $p^{\prime}$), which is called the (geodesic) radius of the neighborhood. Since the points on $\bm $ are equivalent, it suffices to examine the structure of the geodesic neighborhood whose center is located at the point $\{1,1,1\}$.
We can obtain the parametric equation of such a geodesic, if in the equations \bref{geodes} we fix the parameter $s$: $|s|=|R|,$ and changing one of the parameters $q_i$, e.g., $q_1$. In this way, the parametric equation of the unit circle on $\bm $ with the center at the point $\{1,1,1\}$ has the form:
   \begin{equation}\label{edd} X_1=e^{\pm q_1};\quad X_2=e^{\pm q_2},\end{equation}
where the parameter $q_2$ is related to $q_1$ by means of the relation \bref{restr}. Depending on the sign $\pm$ and on the number of branches involved in the dependence \bref{restr}, we obtain for the geodesic neighborhood six connected components. For the case $|R|=1$ we represent one such component in Fig. \ref{ed}.\par
    {\centering\leftskip0em\rightskip5em\small\refstepcounter{figure}\label{ed}
    \includegraphics[width=.5\textwidth, height=0.5\textwidth]{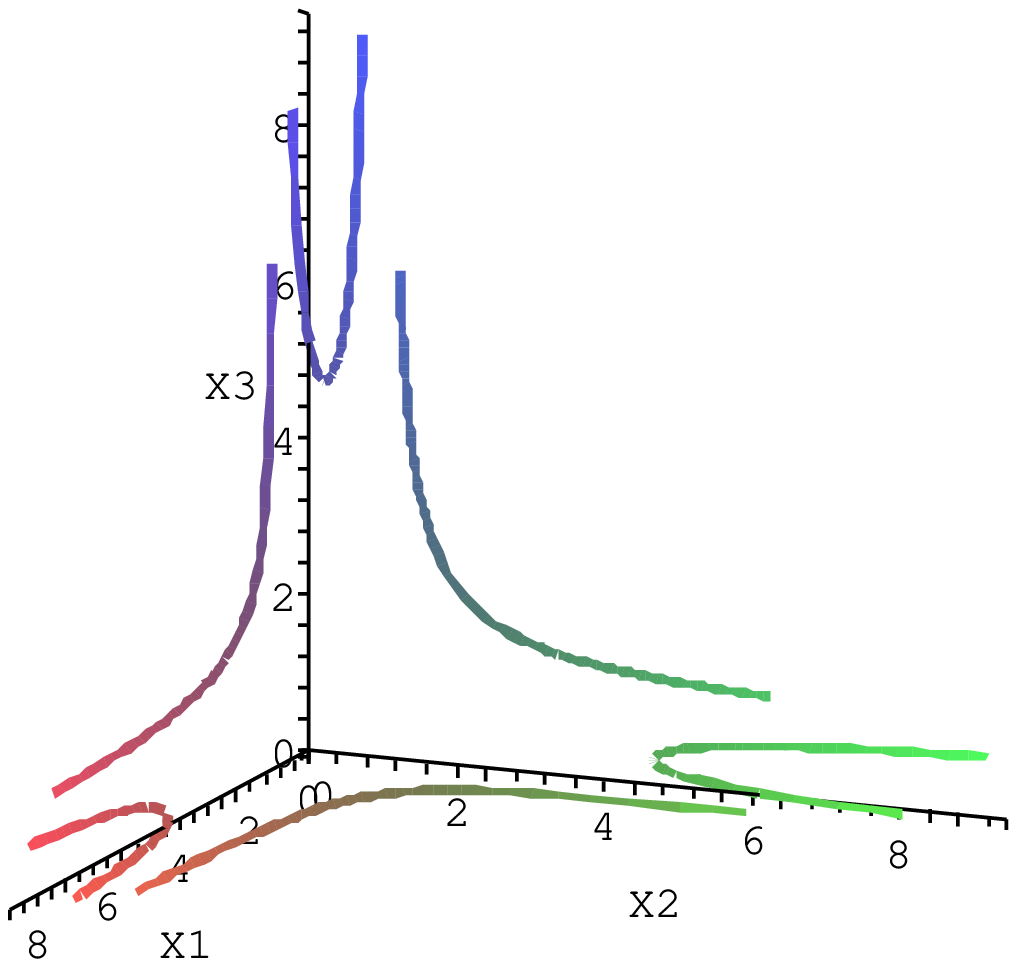}
    \includegraphics[width=.5\textwidth, height=0.5\textwidth]{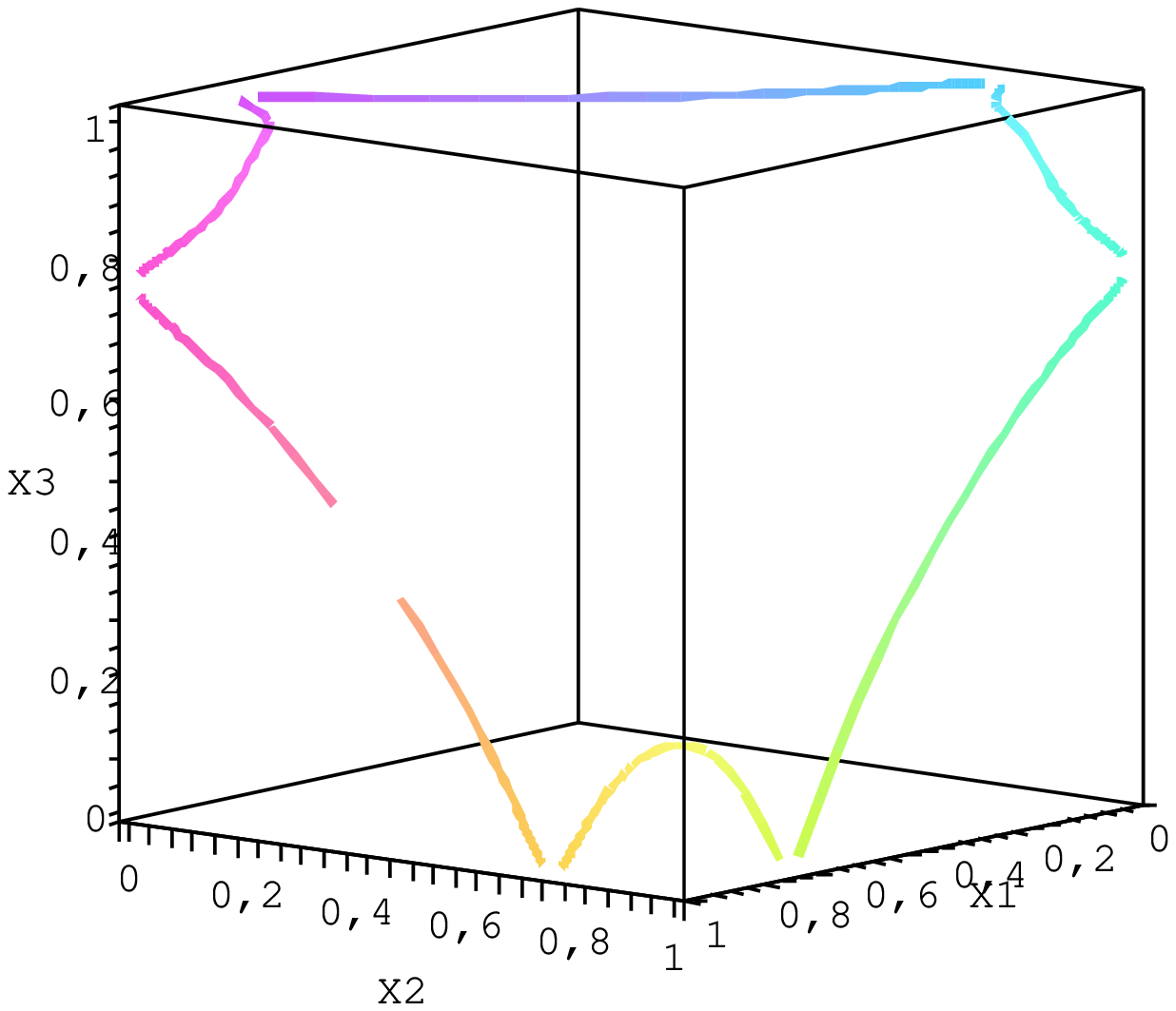}
    \medskip\nopagebreak\par Fig.~\thefigure.
    Six components of the geodesic unit neighborhood. The components of the unit circle with $R=+1$ are the thick
    curves, with $R=-1$ - the thin ones. The thick lines which bend as hyperbolas towards the coordinate planes are
    described by the dependence components \bref{restr} of the corresponding planes. In the right Figure is presented
    the left one (taken with the extension of the borders of the neighborhoods) compactified into a unit cube by
    means of the mapping : $X_i\to\tanh X_i.$}\par\medskip
We shall examine, at last, the question of intersections of geodesics on $\bm $. The parametric equations of pairs of geodesics are:
   \[X_1=a_1e^{q_1s};\quad X_2=a_2e^{q_2s};\quad  \bar X_1=\bar a_1e^{\bar q_1\bar s};\quad \bar X_2=\bar a_2
   e^{\bar q_2\bar s},\]
where the non-barred letters correspond to one geodesic, while the barred ones - to the other one. We find the condition that these intersect as a system of equations:
   \[a_1e^{q_1s}=\bar a_1e^{\bar q_1\bar s};\quad a_2e^{q_2s}=\bar a_2e^{\bar q_2\bar s}.\]
Applying the logarithm and passing the terms containing the parameters $s$ and $\bar s$ to the left-hand-side, we infer a system of linear non-homogeneous equations in terms of these parameters:
   \begin{equation}\label{syspar}q_1s-\bar q_1\bar s=\ln(\bar a_1/a_1);\quad q_2s-\bar q_2\bar s=\ln(\bar a_2/a_2),
   \end{equation}
which define the points of intersections of geodesics on $\bm $. If the determinant of the system \bref{syspar} $\bar q_1q_2-q_1\bar q_2\neq0,$ then the system has a unique solution, and hence, the geodesics intersect at exactly one point. We shall examine now the cases when the determinant of the system \bref{syspar} vanishes. In this situation we have to analyze the system of equations with regards to the parameters of the geodesics:
   \begin{equation}\label{syspar1}\bar q_1q_2-q_1\bar q_2=0;\quad q_1q_2(q_1+q_2)=1;\quad \bar q_1\bar q_2(\bar
   q_1+\bar q_2)=1.\end{equation}
The analysis of this system shows that its unique solution is: $q_1=\bar q_1;\ q_2=\bar q_2$. This means that, if the following condition is {\em not} satisfied:
   \begin{equation}\label{direction}\ln(\bar a_1/a_1)=\ln(\bar a_2/a_2),\end{equation}
the geodesics do not intersect. The condition \bref{direction} in its essence reflects the belonging of the points $(a_1,a_2)$ and $(\bar a_1,\bar a_2)$ to (just) one geodesic. In other words, the following theorem holds true: {\em Through a given point which does not belong to a given geodesic, passes exactly one geodesic which is parallel to the given one.}. A couple of parallel geodesics is illustrated in Fig. \ref{parall}.\par
    {\centering\leftskip0em\rightskip5em\small\refstepcounter{figure}\label{parall}
    \includegraphics[width=.5\textwidth, height=0.5\textwidth]{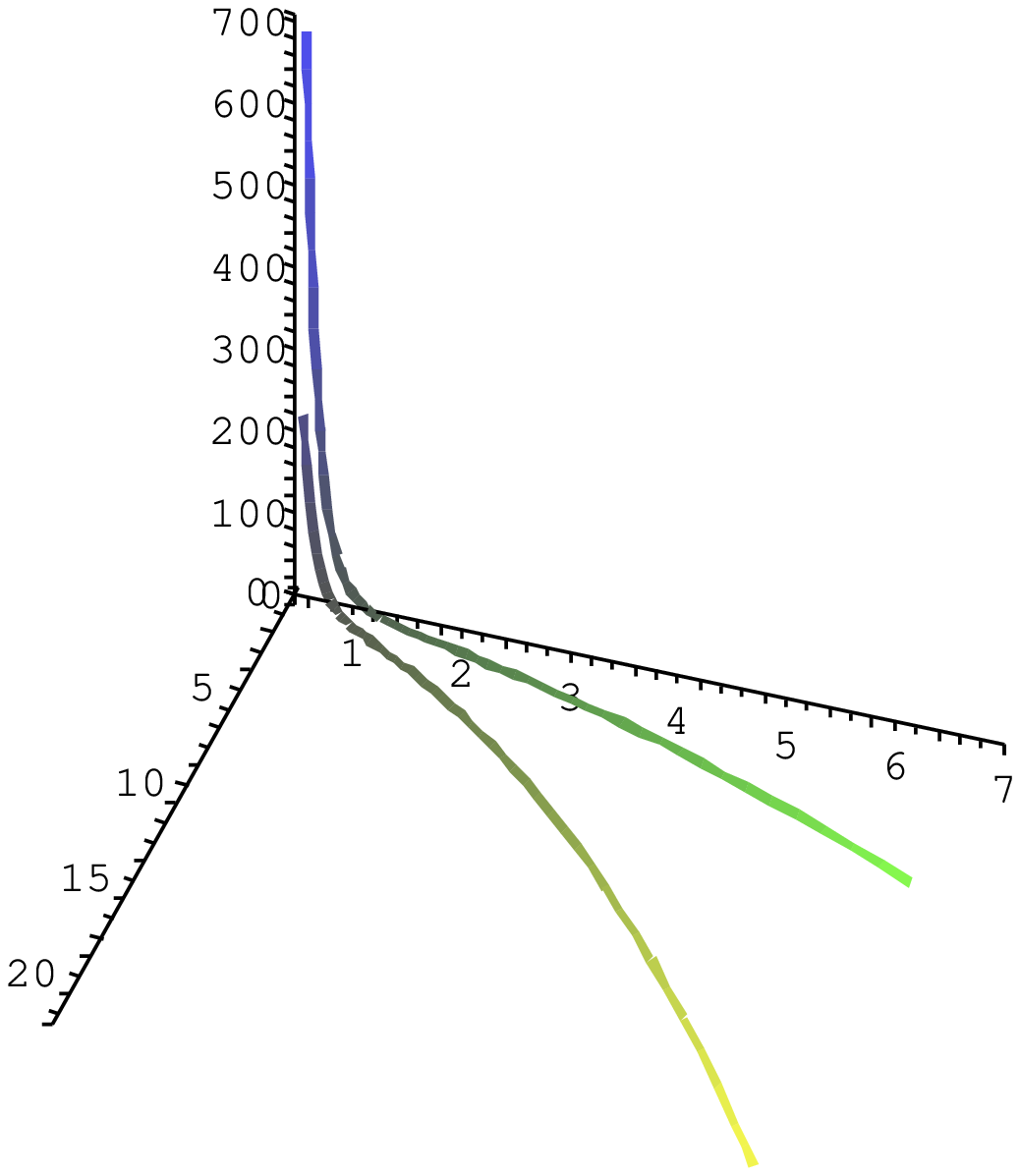}
    \includegraphics[width=.5\textwidth, height=0.5\textwidth]{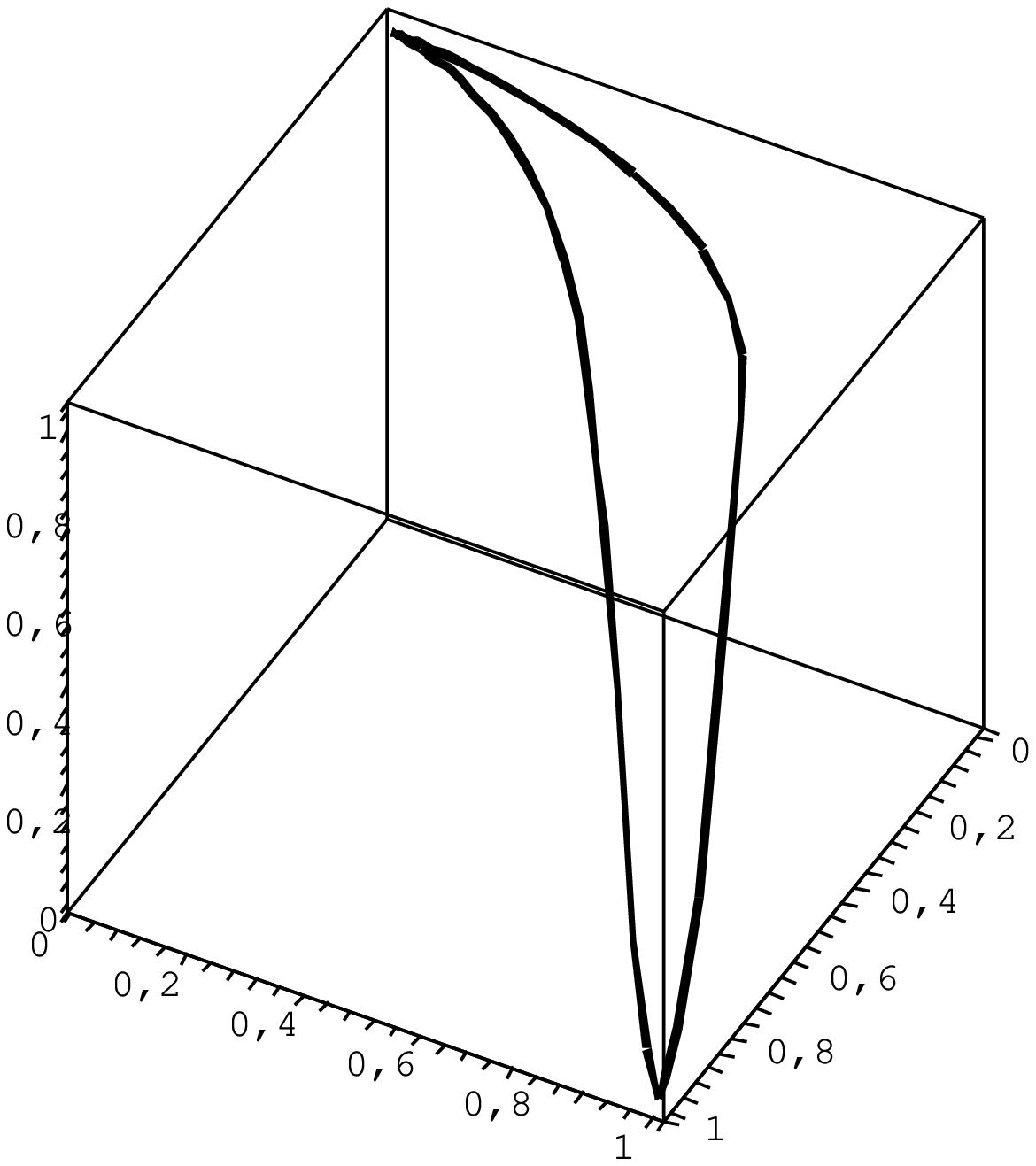}\medskip\nopagebreak\par
    Fig.~\thefigure. A couple of parallel geodesics. In the right image is the left one (taken with extended borders
    of neighborhoods) compactified in the unit cube by means of the mapping: $X_i\to\tanh X_i.$ }\par\medskip
The established property has a similar form to the well-known one from plane Euclidean geometry, expressed in the 5-th Euclid's postulate. The flattening of the indicatrix becomes obvious, if we notice that the induced metric $\mathcal{G}_{S^2_{BM}}$ (the relation \bref{metrS} is mapped to a clear flat form by the change of coordinates: $X_i\to u_i=\ln X_i.$
\section{The explicit expression for bingles}
Using the results of the previous chapter, we shall obtain the explicit expression for additive bingles, according to  \bref{bingl1}. Denoting $\phi[A,B]=s_\ast$, from the definition \bref{bingl1} and the equation \bref{geodes}, we get:
   \begin{equation}\label{eq1}b_1/a_1=e^{q_1s_\ast};\quad b_2/a_2=e^{q_2s_\ast},\end{equation}
where we have fixed $A_1=A_2=1$ considering the initial conditions in the canonic relative to the couple of vectors $A$ and $B$ coordinate system. The formulas \bref{eq1} can be written in the form:
   \[q_1=\frac{1}{s_\ast}\ln(b_1/a_1);\quad q_2=\frac{1}{s_\ast}\ln(b_2/a_2)\]
and replacing these expressions in the normalization conditions \ref{restr}, we obtain the equation:
   \[\frac{1}{s_\ast^3}\ln(b_1/a_1)\ln(b_2/a_2)\ln(b_1b_2/a_1a_2)=1,\]
whence:
   \begin{equation}\label{binged}\phi[a,b]=s_{\ast}=\left[\ln(b_1/a_1)\ln(b_2/a_2)\ln(b_1b_2/a_1a_2)\right]^{1/3}.
   \end{equation}
This formula represents the expression of the additive bingle in terms of the unit vectors, analogously to the Euclidean expression \bref{length2}. Its form, in terms of the components of the initial vectors $A,B$ exhibits
a more symmetric aspect:
   \begin{equation}\label{binggen}\phi[A,B]=\phi[a,b]-\left[\ln\left(\frac{B_1/A_1}{|B|/|A|}\right)\ln
    \left(\frac{B_2/A_2}{|B|/|A|}\right)\ln\left(\frac{B_3/A_3}{|B|/|A|}\right)\right]^{1/3}\end{equation}
   \[\left[\ln\left(\frac{A_1^{2/3}/(A_2A_3)^{1/3}}{B_1^{2/3}/(B_2B_3)^{1/3}}\right)
    \ln\left(\frac{A_2^{2/3}/(A_1A_3)^{1/3}}{B_2^{2/3}/(B_1B_3)^{1/3}}\right)
    \ln\left(\frac{A_3^{2/3}/(A_1A_2)}{B_3^{2/3}/(B_1B_2)^{1/3}}\right)\right]^{1/3}.\]
\section{The Finslerian condition of coplanarity and the operation of bi-conjugation}
As mentioned before, the bingle which was defined in \bref{binggen}, is additive, i.e., for any triple of "coplanar" vectors $A,B,C$ takes lace the following relation, which is similar to \bref{add2}:
   \begin{equation}\label{add3}\phi[A,C]=\phi[A,B]+\phi[B,C].\end{equation}
We have used quotes for the concept of "coplanarity" since it needs to be explained. As it follows from the preceding considerations, from geometric point of view we call coplanar all the vectors whose vertices of the associated unit vectors lie on one of the extremal curves on the indicatrix $\bm$. While displacing a unit vector along such an extremal curve, this vector sweeps some conic surface in $\H$ (using the terminology from \cite{pav1} - "broomshape figures"). We shall call such a conic surface {\it revolution plane}. We have shown above, that the extremal curves, excepting several representatives of their family, are not plane curves in affine sense. This means that {\it the revolution planes and the affine planes are essentially different in the $\H$ geometry}. Moreover, from the point of view of the $\H$ geometry, the affine plane ceases to play any significant role (except the isotropic planes, which are affine planes and which has no intersection points with the metrical ones!).\par
We shall formulate the analytic condition of coplanarity of three vectors $A,B,C$. Passing to the indicatrix, and assuming that the corresponding unit vectors $a,b,c$ lye on the same extremal curve, the following relations hold true:
   \[b_1=a_1e^{q_1s_1};\quad b_2=a_2e^{q_2s_1};\quad c_1=a_1e^{q_1s_2};\quad c_2=a_2e^{q_2s_2},\]
where we have assumed, that the the value $s=0$ corresponds to the position of the end of the vector $a$, the value $s=s_1$ corresponds to the end of the vector $b$ and the value $s=s_2$ corresponds to the end of the vector $c$. Eliminating from the system the parameters $q_1,q_2,s_1,s_2$ of the geodesic, we obtain the conditions of the metric coplanarity of the vectors $A,B,C$ which have the form:
   \begin{equation}\label{comp3}\frac{\ln(b_1/a_1)}{\ln(c_1/a_1)}=\frac{\ln(b_2/a_2)}{\ln(c_2/a_2)}.\end{equation}
Passing from the nit vectors $a,b,c$ to the initial ones $A,B,C,$ this relation can be written in a more expressive way:
   \begin{equation}\label{comp31}\frac{\ln\left(\frac{B_1^{2/3}/(B_2B_3)^{1/3}}{A_1^{2/3}/
    (A_2A_3)^{1/3}}\right)}{\ln\left(\frac{C_1^{2/3}/(C_2C_3)^{1/3}}{A_1^{2/3}/
    (A_2A_3)^{1/3}}\right)}\frac{\ln\left(\frac{B_2^{2/3}/(B_1B_3)^{1/3}}{A_2^{2/3}/(A_1A_3)^{1/3}}
    \right)}{\ln\left(\frac{C_2^{2/3}/(C_1C_3)^{1/3}}{A_2^{2/3}/(A_1A_3)^{1/3}}\right)}\end{equation}
The existence of the third coordinate in this expression is evidently accidental and is connected to the fact that the coordinate chart used for describing the indicatrix $\bm $ and its geodesics was related to the plane $\{X_1,X_2\}.$ The description of the same metric plane in different charts would have completed the relation \bref{comp31} with a new equation, which completely rounds up the symmetry of coordinates and vectors. The full condition of metric coplanarity has the form:
   \begin{equation}\label{comp32}\frac{\ln\left(\frac{B_1^{2/3}/(B_2B_3)^{1/3}}{A_1^{2/3}/
   (A_2A_3)^{1/3}}\right)}{\ln\left(\frac{C_1^{2/3}/(C_2C_3)^{1/3}}{A_1^{2/3}/(A_2A_3)^{1/3}}
    \right)}\frac{\ln\left(\frac{B_2^{2/3}/(B_1B_3)^{1/3}}{A_2^{2/3}/(A_1A_3)^{1/3}}
    \right)}{\ln\left(\frac{C_2^{2/3}/(C_1C_3)^{1/3}}{A_2^{2/3}/(A_1A_3)^{1/3}}\right)}
    \frac{\ln\left(\frac{B_3^{2/3}/(B_1B_2)^{1/3}}{A_3^{2/3}/(A_1A_2)^{1/3}}\right)}{\ln
    \left(\frac{C_3^{2/3}/(C_1C_2)^{1/3}}{A_3^{2/3}/(A_1A_2)^{1/3}}\right)}\end{equation}
and in this form it is, obviously, equivalent to the Euclidean coplanarity condition \bref{compp}. To prove this non-accidental analogy, we shall show that there exists as well a Finslerian analogue of the more compact condition \bref{compc}. For this we define the mapping $\flat$: $\H \to\mathcal{H}^{\flat}_3$, which acts according to the rule:
   \begin{equation}\label{log}A=\{A_1,A_2,A_3\}\mapsto A^{\flat}=\{\ln(A_1^{2/3}/(A_2A_3)^{1/3}),
    \ln(A_2^{2/3}/(A_1A_3)^{1/3}),\ln(A_3^{2/3}/(A_1A_2)^{1/3}).\end{equation}
We shall call this mapping {\it bi-projection of} $\H$, the space $\H^{\flat}$ as {\it bi-space} over  $\H$, and the element $A^{\flat}$ - {\it the bingle} of the element $A$. We note that the bi-space $\H^{\flat}$ is 2-dimensional, as consequence of the identically satisfied relation:
   \begin{equation}\label{trace}\text{Tr}\,A^{\flat}\equiv A_1^{\flat}+A_2^{\flat}+A_3^{\flat}=0.\end{equation}
We remark as well, that the bi-projection is non-linear: $(A+B)^{\flat}\neq A^{\flat}+B^{\flat}.$\par
It is easy to check that the first equality \bref{comp32} is practically the 12-th component of the more compactly written relation:
   \begin{equation}\label{comp3f}(A^{\flat}-B^{\flat})\wedge(A^{\flat}-C^{\flat})=0,\end{equation}
while the second equality is the 13-th component of this relation. Accordingly, we obtain the 23-rd component of this relation if we examine the equality of the first and third power in \bref{comp32}. In the Euclidean and in the general affine space, a relation of the form \bref{comp3f} means exactly the fact that the points having the position vectors  $A^{\flat},B^{\flat},C^{\flat}$ belong to the same affine line (in the 3-dimensional case, $\wedge$ is the cross-product $\times$). \par
Hence, we draw two main conclusions:\par
1) {\it the Euclidean condition \bref{compc} of vector coplanarity for which the additivity condition holds true, has a Finslerian-hyperbolic analogue - the condition \bref{comp3f} of collinearity for the bingles $A^{\flat}-B^{\flat}$ and $B^{\flat}-C^{\flat}$ or the collinearity of points $A^\flat, B^\flat, C^\flat$ i $\H^\flat$};\par
2)  {\it to any plane of revolution in  $\H$ corresponds some affine line in $\H^{\flat}$ and converse: to each affine line in $\H^{\flat}$ there exists a plane of revolution in $\H$.}\par
We note, that the obtained result confirms and sets the rightness of the hypothesis of the "nonlinear coplanarity condition", formulated earlier in \cite{kok}. Fig. 7 clearly illustrates the difference between the plane of revolution and the affine plane in $\H$.\par
\begin{center}
    {\centering\leftskip5em\rightskip5em\small\refstepcounter{figure}\label{paralld}
    \includegraphics[width=.5\textwidth, height=0.5\textwidth]{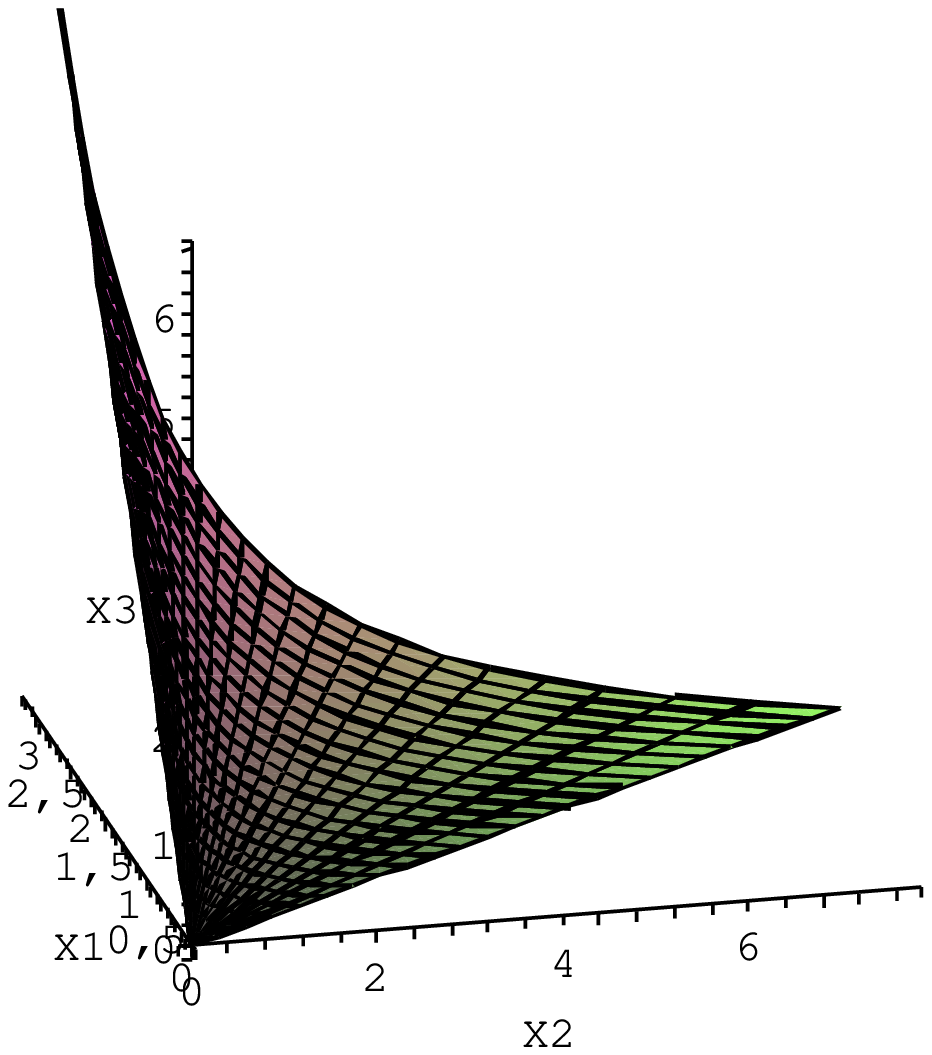}\medskip
    \nopagebreak\par Fig.~\thefigure. {\small A fragment of the metric plane of revolution}}
\end{center}
\section{The geometric properties of the space $\H^{\flat}$}
The bi-projection mapping has a deeper fundamental character, than representing the means of setting the formal analogy between the conditions of affine and metric coplanarity. Indeed, in terms of vectors of the space $\H^{\flat}$, the expression \bref{binggen} of a bingle can be rewritten in the following unexpectedly simple form:
   \begin{equation}\label{dual}\phi[A,B]=|A^{\flat}-B^{\flat}|,\end{equation}
where the norm in the space $\H^{\flat}$ is given by the formula \bref{norm}, assuming that in this space is defined the Berwald-Moor metric of the form \bref{bm}. This formula and its consequences have a deep geometric character. Before discussing them, we shall investigate in detail the geometric aspects of the space $\H^{\flat}$ and of the bi-projection mapping - whose image this space is. As mentioned before, the maping $\flat$: $\H\to\H^{\flat}$ transforms the 3-dimensional linear space $\H$ into the 2-dimensional manifold $\H^{\flat}$, which is in fact a 2-dimensional linear space. Indeed, the fundamental property of its points - \bref{trace}, i.e. the vanishing of traces, is an invariant relative to taking linear combinations:
   \[\text{Tr}(\lambda A^{\flat}+\mu B^{\flat})=0,\quad \text{если}\quad\text{Tr}A^{\flat}=\text{Tr}B^{\flat}=0.\]
This means that $\H^{\flat}$ is a linear space, and $\text{dim}\H ^{\flat}=2.$ This space can be represented by means of its embedding in the 3-dimensional linear space $\Omega\H$, which is set in the same manner as the initial one, $\H$ and which is endowed - according to \bref{dual}, with the standard Berwald-Moor metric. Such an embedding in $\Omega\H$ is represented as a plane, passing through the origin, and which is orthogonal (in the Euclidean sense) to the vector $I=\{1,1,1\}$ (cf. Fig. \bref{embed}). The equation of this plane is given by the vanishing of the sum of coordinates.\par
    \begin{figure}[htb]\centering \unitlength=0.50mm \special{em:linewidth 0.4pt}
    \linethickness{0.4pt} \footnotesize \unitlength=0.4mm
    \special{em:linewidth 0.4pt} \linethickness{0.4pt}\input{two-15.pic}
    \caption{\small The embedding of $\H^\flat$ in $\Omega\H$.}\label{embed}\end{figure}\par
Hence, the bi-projection mapping transforms vectors from $\H$ into vectors of the plane $\H^{\flat}\subset\Omega\H$. We note, that the bi-projection mapping represents an interesting geometric example of nonlinear transformation between vector spaces. Due to the nonlinearity, this mapping cannot be the subject of most of the standard theorems regarding morphisms of vector spaces.\par
Like any projection, the bi-projection is onto\footnote{We remind, that a surjective map for which each element of the range has a pre-image.} as mapping $\H\to\H^{\flat}$ (but is not onto as mapping $\H\to\Omega\H$). We shall find the $A^{\flat}$-fiber of the bi-projection, i.e., the set of elements $X\in\H$, for which $X^{\flat}=A^{\flat}$. To this aim, we note that any two vectors of $\H$ which differ only by their norm, are taken by the bi-projection into the same element of $\H$. We shall call the subsets of $\H$ of the form $\cup_{\lambda\in \R}\lambda A\equiv\ell(A)$ as
{\it the rays having the direction $A$}. In this way, any two points on a ray are joined by the bi-projection into a single point of the plane $\H^{\flat}.$\par
Since each ray in $\H$ is uniquely determined by the unit vector of the direction: $\ell(A)=\ell(a)$, the construction of rays can be considered on the unit sphere, on which the antipodal points $a$ and $-a$ are identified. Such a sphere will be called {\it the projective unit sphere Berwald-Moor} and will be denoted by $P\bm$. The projective sphere contains 4 connected components and is obtained from the sphere $\bm $ by identifying the antipodal points. For any element of the projective sphere having the coordinates $\{a_1,a_2,1/(a_1a_2)\}$ the general formulas of the bi-projection \bref{log} get the form:
   \[a^{\flat}=\{\ln a_1,\ln a_2,-\ln(a_1a_2)\},\]
from which it follows that on each component of $P\bm$ the bi-projection acts bijectively. This means that {\it the fibers of the bi-projection are exactly the rays of the space $\H$.}\par
We note that the elements $\Omega\H$ which do not belong to $\H^{\flat}$ do not have pre-images in $\H$ and cannot be viewed as bingles. As well, we remark that the "kernel" of the bi-projection is the ray $\ell(I),$ where $I=\{1,1,1\},$ since $I^{\flat}=0$ and if $X^{\flat}=0,$ then $X=I.$\par
We shall further study the symmetries of $\H^{\flat}.$ Since $\H^{\flat}$ is a linear space, the translations $\mathcal{T}\H^{\flat}$ and the multiplication with real numbers $\mathcal{D}\H^{\flat}$ leave the space of bingles invariant. Geometrically, this mapping describes the sliding of the plane $\H^{\flat}$ along itself in $\Omega\H$, and its homogeneous scaling. We shall now examine the nonlinear transformations $\mathcal{N}\H^{\flat}$ which transform $\H^{\flat}$ into itself. These mappings are described by the following formulas:
   \begin{equation}\label{nonl}A^{\flat}\to A^{\flat}_{\lambda}=\{\lambda_1A_1^{\flat},
   \lambda_2A_2^{\flat},\lambda_3A_3^{\flat}\},\end{equation}
where the transformation vector $\lambda\in\Omega\H$ lays in the plane, which in Euclidean sense is orthogonal to $A^{\flat}$. Indeed, the Euclidean orthogonality of the vectors $A^{\flat}$ and $\lambda$ is equivalent to the condition: $\text{Tr}A^{\flat}_{\lambda}=\lambda_1A_1^{\flat}+\lambda_2A_2^{\flat}+\lambda_3A_3^{\flat}=0$. This means that the transformed vector $A^{\flat}_{\lambda}$ is a bingle. Since the transformation depends on the vector, it is non-linear. We note, that the elements of the mapping are in general vectors from $\Omega\H$, i.e., the vectors of this space can be regarded as elements of the set of outer automorphisms of the bingle space. This set contains the unit $I=\{1,1,1\}$ (this vector is orthogonal in Euclidean sense to all the vectors from $\H^{\flat}$, hence it is applicable to all vectors, and here each vector is mapped into itself); for the case, when all $\lambda_i\neq0,$ the mapping is invertible: $(A^{\flat}_{\lambda})_{\lambda^{-1}}=A^{\flat},$ where $\lambda^{-1}=\{\lambda_1^{-1},\lambda_2^{-1},\lambda_3^{-1}\}$ and the composition of mappings in the case when $A^\flat_\lambda\perp\sigma$ has the form:
   \[(A^{\flat}_{\lambda})_{\sigma}=A^{\flat}_{\lambda\sigma},\]
where $\lambda\sigma=\{\lambda_1\sigma_1,\lambda_2\sigma_2,\lambda_3\sigma_3\}.$ We can say, that the mapping from
$\mathcal{N}\H ^{\flat}$ generate a partial algebra, which is a partial subalgebra of the algebra of poly-numbers $P_3.$\par
The formula \bref{dual} means that in the space $\H^{\flat}$ is defined the Berwald-Moor metric. Besides the mentioned before translations, its isometries are described by hyperbolic rotations $\mathcal{D}_2^{\flat},$ which act according to the rule \bref{nonl}, in which instead the limitation related to the Euclidean orthogonality, is imposed the condition: $\lambda_1\lambda_2\lambda_3=1$, i.e., the vertex of the vector $\lambda$ has to belong to he unit sphere $\bm$ in $\Omega\H^{\flat}.$ The intersection $\mathcal{N}\H^{\flat}\cap\mathcal{D}_2^{\flat}\equiv\mathcal{N}
\mathcal{D}^{\flat}$ generates an 1-parametric family of nonlinear isometries $\H^{\flat}$. Geometrically, the vectors of the mappings $\lambda\in\mathcal{N}\mathcal{D}^{\flat}$ have their vertices laying on a hyperbola, which is the intersection of the unit sphere $\bm$ and the plane which is orthogonal to the vector $A^{\flat},$ on which this mapping acts.
\section{Further properties of bingles}
Based on the results of the previous section, we continue the study of the most important properties of bingles. In order to make a distinction between bingles  $\phi[A,B]$ between vectors and bingles as elements of the space $\H^{\flat},$ we shall call the bingles of the form $\phi[A,B]$ as {\it reciprocal bingles.}
   \begin{enumerate}
\item The formula \bref{dual} means, that the space $\H ^{\flat}$ is a metric space, whose metric is induced in a non-trivial way from the metric of $\H $ by means of bi-projection. In fact, this metric is - according to its form, the Berwald-Moor metric.
\item The space $\H ^{\flat}$ itself consists of objects (points) of a new geometric nature: according to the formula \bref{dual} {\it the distance between two points of this space has the meaning of angle between the corresponding pre-image vectors from $\H$.} Such a duality - in the framework of quadratic geometry has been discussed by P.K. Rashevsky in \cite{rush}. A significant factor which appears in the geometry of poly-numbers, is {\it the coincidence of metrics in the space of vectors, and angles formed by these vectors.}
\item We note, that the bi-projection lives beyond the framework of conformal, and even analytic (in poly-numbers sense) transformations \cite{gar,gard}.
\item It is easy to remark, that the elements of the space $\H ^{\flat}$ - from the point of view of the poly-numbers algebra, represent exactly the exponential angles of the poly-numbers:
   \[A^{\flat}_1=\chi_1;\quad A^{\flat}_2=\chi_2;\quad A^{\flat}_3=\chi_3,\]
where $\chi_i$ are the exponential angles \bref{appare}. In fact, the formula \bref{dual} can be re-written in terms of exponential angles, as follows:
   \begin{equation}\label{expform}\phi[A,B]=[(\chi_1^A-\chi_1^B)(\chi_2^A-\chi_2^B)(\chi_3^A-\chi_3^B)]^{1/3}.
   \end{equation}
\item The bi-projection mapping provides an isomorphism between the general group of homotheties $\mathcal{D}_2$ and $\H $ the group of translations $\mathcal{T}_2^{\flat}$ в $\H ^{\flat}$:
   \begin{equation}\label{hom}A=\{A_1,A_2,A_3\}\mapsto\mathcal{D}_{\alpha_1,\alpha_2\alpha_3}A=
    \{\alpha_1A_1,\alpha_2A_2,\alpha_3A_3\}\stackrel{\flat,\flat^{-1}}\rightleftarrows\end{equation}
   \[A^{\flat}=\{A_1^{\flat},A_2^{\flat},A_3^{\flat}\}\mapsto\mathcal{T}_{\tau_1,\tau_2,\tau_3}
    A^{\flat}\{A_1^{\flat}+\tau_1,A_2^{\flat}+\tau_2,A_3^{\flat}+\tau_3\},\]
where
   \[\aru{\tau_1=\frac{2}{3}\ln\alpha_1-\frac{1}{3}\ln\alpha_2-\frac{1}{3}\ln\alpha_3;\mm
    \tau_2=\frac{2}{3}\ln\alpha_2-\frac{1}{3}\ln\alpha_1-\frac{1}{3}\ln\alpha_3;\mm
    \tau_3=\frac{2}{3}\ln\alpha_3-\frac{1}{3}\ln\alpha_1-\frac{1}{3}\ln\alpha_2.}\]
Hence, {\it the translational invariance of reciprocal bingles represents the $\flat$-image of their conformal invariance.}
\item We examine the reciprocal bingle of the form $\phi[I,A].$ According to \bref{dual}, it is equal to $|A^{\flat}|\equiv(\chi^A_1\chi^A_2\chi^A_3)^{1/3},$ where $\chi_i^A$ are the exponential angles of the poly-number  $A$. Such a bingle measures the deviation of the direction $A$ from the direction of the unit, which geometrically coincides with the spacial bisector of the first coordinate octant. Similar constructions of the form $\phi[I_{(j)}A],$ where $I_{(j)}$ is the spacial bisector of the $j$-th coordinate octant, allow us to extend the definition of reciprocal bingles between vectors, to other octants. We note, that the reciprocal bingle between vectors, which lay in different octants will be necessarily complex.
\item In the figure \ref{ill} are presented illustrative diagrams of the reciprocal bingle $\phi[I,A]$ in the positive octant.
\end{enumerate}
    {\centering\leftskip0em\rightskip5em\small\refstepcounter{figure}\label{ill}
    \includegraphics[width=.5\textwidth, height=0.5\textwidth]{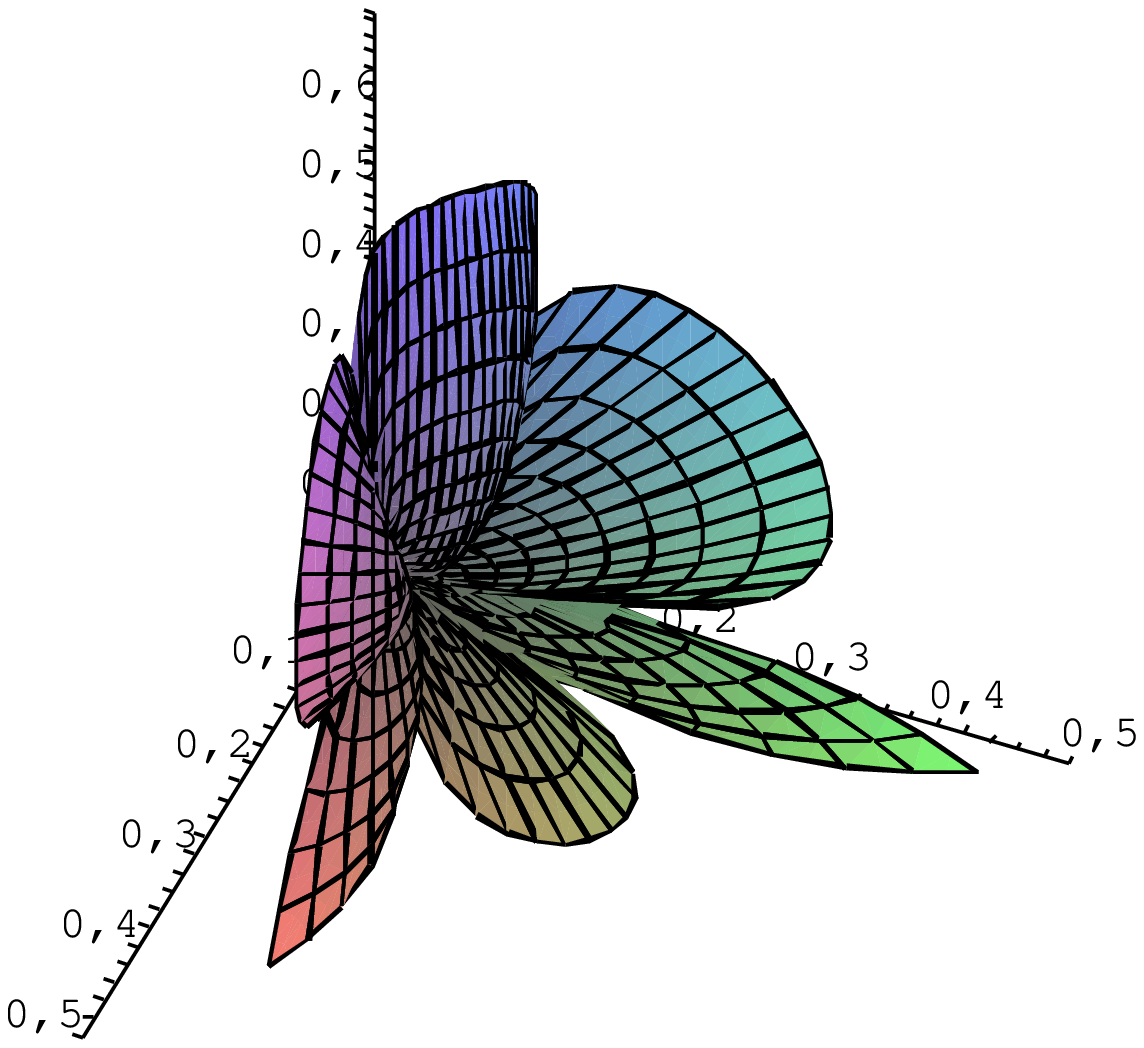}
    \includegraphics[width=.5\textwidth, height=0.5\textwidth]{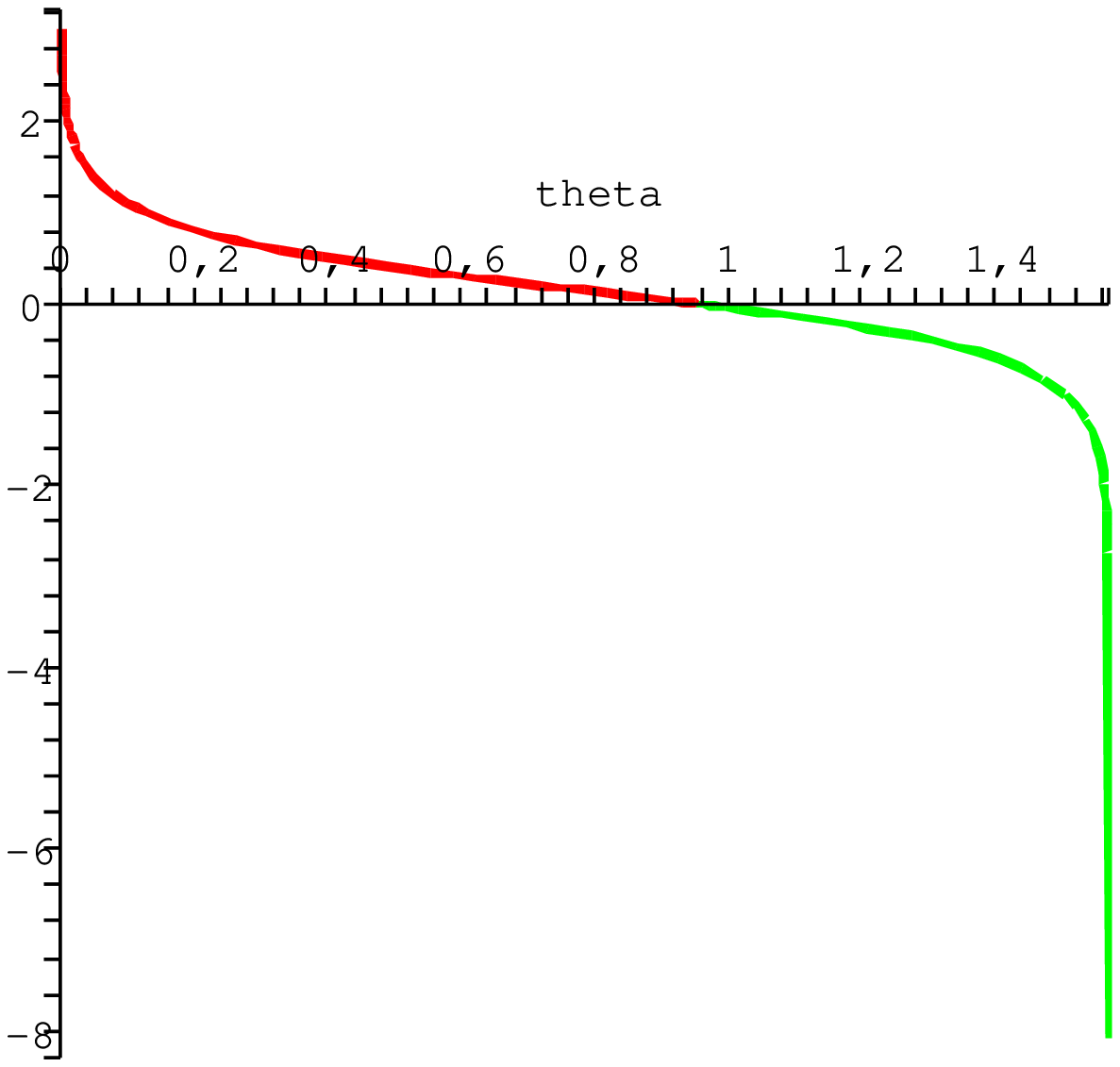}\medskip\nopagebreak\par
    Fig.~\thefigure. In the left image is represented the spatial Euclidean chart of the reciprocal bingle
    $\phi[I,A]$: to each direction $A$ in $\H $ the surface provides correspondingly the value of the modulus of the
    bingle $\phi[I,A]$. Numerically, this is equal to the Euclidean distance from the origin
    of coordinates to the point of intersection between the ray $\ell(A)$ with the surface. In the right figure, is
    constructed the graph of the dependence of the reciprocal bingle $\phi[I,A]$ - where the vector
    $A=(\alpha\sin\theta,\alpha\sin\theta,\alpha\cos\theta),$ and $\theta$ is the standard spherical angle, as a
    function of $\theta.$}\par\medskip
\section{The second (relative) bingle}
The definition of the second independent bingle can be stated by analogy to the one of the azimuthal angle $\varphi$ of the standard angular system of coordinates on the Euclidean sphere. The first bingle becomes, in this case, the hyperbolic analogue of the latitude angle $\theta.$ Like in the Euclidean case, the second bingle needs to have a prior fixed (arbitrary) direction of reference. In order to distinguish this bingle from the elements $\H ^{\flat}$ and the reciprocal bingle, we shall call it {\it relative bingle}.\par
As a first step in defining the second bingle, we shall transform a pair of arbitrary unit vectors $a$ and $b$ on the unit sphere to their canonic position, such that $a=\{1,1,1\},$ $b=\{b_1/a_1,b_2/a_2,b_3/a_3\}$ and we shall find the point of intersection of the geodesic arc which joins $a$ and $b$, with the geosdesic of the unit circle whose center is $a$. The corresponding system of geodesic equations for the geodesic arc are:
   \[X_1=e^{q_1s};\quad X_2=e^{q_2s},\]
where
   \[q_1=\frac{1}{s_\ast}\ln(b_1/a_1),\quad s_\ast=\phi[a,b],\quad q_1q_2(q_1+q_2)=1,\]
and for the unit circle,
   \begin{equation}\label{edcirc}Y_1=e^{\bar q_1};\quad Y_2=e^{\bar q_2},\quad \bar q_1\bar q_2(\bar q_1+
   \bar q_2)=1.\end{equation}
After eliminating the parameter $s,$ we get the equation: $q_1\bar q_2-q_2\bar q_1=0,$ whose shape resembles the first equation \bref{syspar1}. Its solution is already known:: $q_1=\bar q_1,$ $q_2=\bar q_2.$\par
As second step, we remark that, due to the 16-connectedness of the unit sphere, we should distinguish the cases when the beforementioned intersection point between the geodesic arc and the unit circle falls into its different components. As shown before in Section \bref{extr}), the components of the circle which contain the direction
$a$ --- $b$ can be identified within our parametrization by means of the signs of the parameters:
   \begin{equation}\label{formul1}q_{1}[a,b]=\frac{1}{s_\ast}(B_1^{\flat}-A_1^{\flat})=\frac{B_1^{\flat}-
   A_1^{\flat}}{|A^{\flat}-B^{\flat}|};\quad q_{2}[a,b]=\frac{1}{s_\ast}(B_2^{\flat}-A_2^{\flat})=
   \frac{B_2^{\flat}-A_2^{\flat}}{|A^{\flat}-B^{\flat}|}\end{equation}
(these expressions are the hyperbolic analogues of the director cosines of the vectors in the Euclidean space) and of the sign of the bingle itself
   \[s_\ast=|A^{\flat}-B^{\flat}|.\]
For definiteness, we assume in the case $q_1>0,q_2>0$ and $s_\ast\lessgtr0$ we use the parametrization in the plane $(X_1X_2)$ (the third positive and negative component is $3^{\pm}$); in the case $q_1<0,q_2>0$ and $s_\ast\lessgtr0$ we use the parametrization in the plane $(X_1X_3)$ (the second positive and negative component is $2^{\pm}$); at last, in the case $q_1>0,q_2<0$ and $s_\ast\lessgtr0$ we use the parametrization in the plane $(X_2X_3)$ (the first positive and negative component is $1^{\pm}$). Then, in hte first case the parametrization of the arc of the unit circle has the form:
   \[X_1=e^{\pm q_1};\quad X_2=e^{\pm q_2};\quad q_1q_2(q_1+q_2)=1,\]
in the second:
   \[X_1=e^{\pm q_1};\quad X_3=e^{\pm q_3};\quad q_1q_3(q_1+q_3)=1,\]
and in the third
   \[X_2=e^{\pm q_2};\quad X_3=e^{\pm q_3};\quad q_2q_3(q_2+q_3)=1.\]
The common - for all the components, direction of displacement along the unit circle (which can be represented as a connected curve in the compactified $\H $ (the second graphic in Fig. \ref{ed}), is determined by the following rules of change for parameters (we start with the third positive component whose direction is taken from $X_2$ to$X_1$):
   \[3^{+}:\ q_1\in(0;\infty)\to 1^{-}:\ q_2\in(0;\infty)\to 2^{+}:\ q_3\in(0;\infty)\to\]
   \[ 3^{-}:\ q_1\in(0;\infty)\to 1^{+}:\ q_2\in(0;\infty)\to2^{-}:\ q_3\in(0;\infty).\]
Now we are able to define the second bingle $\psi[a,b]$ between $a$ and $b$ as the length of the arc of the unit circle, enclosed between some distinguished point of this circle and the determined above point of its intersection with the geodesic arc which joins $a$ with $b.$ Depending on the connected component in which this point is located,, we obtain a distinct relative bingle. For this reason, the relative bingle $\psi[a,b]$ can be endowed with supplementary indices: $\psi_j^{\pm}[a,b]$, which show to which connected component this bingle relates. For instance, the notation $\psi_1^+[a,b]$ shows, that the bingle relates to the first positive connected component of the unit circle, etc. As distinguished point taken as origin of coordinates in each connected component, we shall choose "the symmetric point": in the first component to this point correspond the values $q_2=q_3=2^{-1/3},$ in the second: $q_1=q_3=2^{-1/3},$ in the third $q_1=q_2=2^{-1/3}$. Then the formulas for bingles on different connected components of the unit circle get the following form:
   \begin{equation}\label{bbingle2}\psi_1[a,b]=\int\limits_{2^{-1/3}}^{q_2[a,b]}\frac{ds}{dq_2}dq_2;\quad
   \psi_2[a,b]=\int\limits_{2^{-1/3}}^{q_3[a,b]}\frac{ds}{dq_3}dq_3;\quad
    \psi_3[a,b]=\int\limits_{2^{-1/3}}^{q_1[a,b]}\frac{ds}{dq_1}dq_1,\end{equation}
where the integration is taken along the arcs of the corresponding components.\par
Due to the symmetry of all the components, it suffices to determine the explicit form of the integrals for one of them only. We shall examine in detail the integral for the third component. Replacing in the formula \bref{func} the parametric dependence of the form \bref{edcirc}, we get after several elementary simplifying calculations:
   \[\psi_3[a,b]=\int\limits_{2^{-1/3}}^{q_1[a,b]}(\dot q_2+\dot q_2^2)^{1/3}\,dq_1,\]
where the dot represents the differentiation of the parameter $q_2$ relative to $q_1$, taking into account their functional dependence provided by the last equation in \bref{edcirc}. Using the differential consequence of this equation \bref{edcirc}:
   \begin{equation}\label{diffseq}\frac{dq_2}{dq_1}=-\frac{q_2(q_2+2q_1)}{q_1(q_1+2q_2)}\end{equation}
and its explicit solution on its principal branch:
   \[q_2=\frac{\sqrt{q_1^4+4q_1}-q_1^2}{2q_1},\]
we obtain, after performing several transformations, to the needed expression for the arc-length on the third component of the unit circle: $\psi_3[a,b]=F(q_1[a,b]),$ where the function $F(\xi)$ is given by the following integral:
   \begin{equation}\label{Fbing}F(\xi)\equiv-\frac{1}{2}\int\limits_{2^{-1/3}}^{\xi}\left(\frac{(x^2-
    \sqrt{x(x^3+4)})(3x^2+\sqrt{x(x^3+4)})\sqrt{x(x^3+4)}+x^3-2)}{x^4(x^3+4)}\right)^{1/3}\,dx.\end{equation}
The "minus" sign in front of the integral has the role that, while following the unit circle in the chosen by us positive sense (on the second Fig. \ref{ed} to it corresponds the displacement along a "closed curve" in clockwise sense), the bingle increases. The rightness of such a choice of sign can be explained by the form of dependence of the cube of the integrated function from the expression \bref{Fbing}, which is represented in Fig. \ref{podint}.\par
    {\centering\leftskip5em\ \rightskip5em\small \refstepcounter{figure}\label{podint}
    \includegraphics[width=.5\textwidth, height=0.5\textwidth]{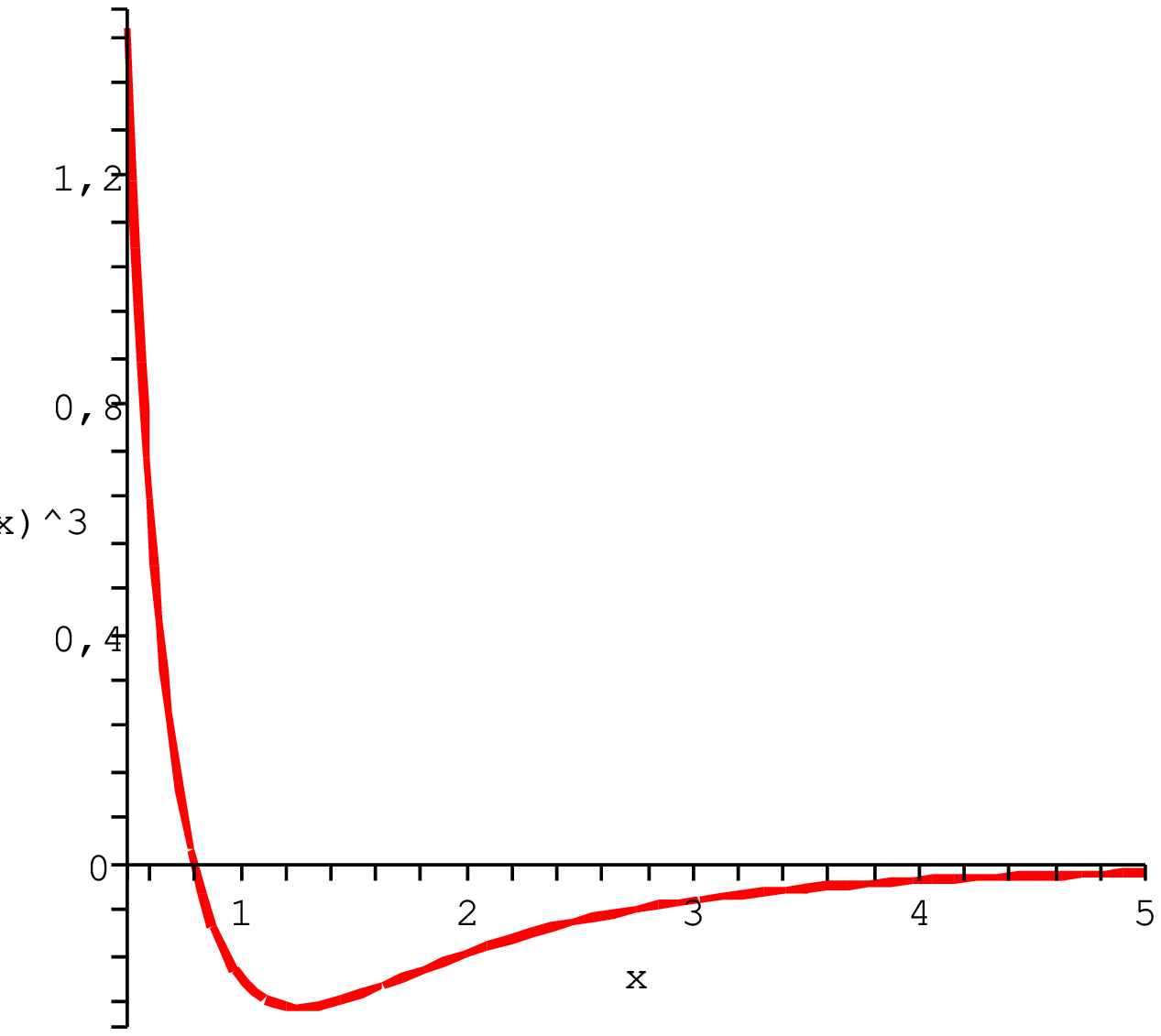}\medskip\nopagebreak\par
    Fig.~\thefigure. The form of the function $F^{\prime}{}^3.$ This function changes its sign at the point
    $x=2^{-1/3}.$}\par\medskip
We write the final expressions for the second bingle, taking into account the formulas \bref{formul1} and \bref{bbingle2}:
   \begin{equation}\label{bbingf}\psi_1[A,B]=F\left[\frac{B_1^{\flat}-A_1^{\flat}}{|A^{\flat}-
    B^{\flat}|}\right];\quad \psi_2[A,B]=F\left[\frac{B_2^{\flat}-A_2^{\flat}}{|A^{\flat}-B^{\flat}|}\right];\quad
    \psi_3[A,B]=F\left[\frac{B_3^{\flat}-A_3^{\flat}}{|A^{\flat}-B^{\flat}|}\right],\end{equation}
where the function $F$ is determined by the formula \bref{Fbing}. The function $F,$ which cannot be expressed in terms of elementary functions, ensures the additivity of the defined bingle (see \cite{pav2}).
\section{The properties of the second bingle}
Having in view the before mentioned analogy of the arguments of the function $F$ in formulas \bref{bbingf} with the director cosines of vectors in the Euclidean space, the function $F$ should be considered as the Finslerian-hyperbolic analogue of the function $\arccos,$ and its inverse, $F^{-1}\equiv\text{cfh}$, the Finslerian-hyperbolic analogue of the cosine. Thus, we have\footnote{The minus sign in the definition of the function $\text{cfh}$ is introduced for the sake of convenience.}:
   \begin{equation}\label{cfh}
   \frac{B_1^{\flat}-A_1^{\flat}}{|A^{\flat}-B^{\flat}|}\equiv-\text{cfh}(\psi_1[A,B]);\quad
   \frac{B_2^{\flat}-A_2^{\flat}}{|A^{\flat}-B^{\flat}|}\equiv-\text{cfh}(\psi_2[A,B]);\quad
   \frac{B_3^{\flat}-A_3^{\flat}}{|A^{\flat}-B^{\flat}|}\equiv-\text{cfh}(\psi_3[A,B]).\end{equation}
Then we have {\it the fundamental identity of Finslerian trigonometry:}
   \begin{equation}\label{identi}\text{cfh}\,\psi_1\text{cfh}\,\psi_2\text{cfh}\,\psi_3=1.\end{equation}
This relation has the same meaning as the Euclidean relation of normalized director cosines:
   \[\cos^2\alpha+\cos^2\beta+\cos^2\gamma=1,\]
where $\alpha,\beta,\gamma$ are the angles between the vector and the axes of he Cartesian system of coordinates. By comparing the identity \bref{identi} and the condition \bref{normir} imposed on the parameters of the group $\mathcal{D}_2,$ we are led to the conclusion thet the parameters of the group can have the meaning of the cosines, which parametrize the transformations of the group $\mathcal{D}_2.$ Indeed, as it can be seen from formula \bref{cfh}, to each vector $A\in\H ,$ corresponds the transformation $D_{\text{cfh}\psi_1,\text{cfh}\psi_2,\text{cfh}\psi_3}\in\mathcal{D}_2,$ where $\psi$ is the relative bingle between $A$ and $I=\{1,1,1\}.$ Hence, the functions $\text{cfh}_i$ provide a natural parametrization for the hyperbolic rotations, analogous to the Euler angles from the Euclidean geometry.\par
We can introduce the Finslerian-hyperbolic analogues of sines, tangents and co-tangents, by means of the formulas:
   \[\text{sfh}\,\psi_1=\text{cfh}\,\psi_2\text{cfh}\,\psi_3;\quad\text{sfh}\,\psi_2=\text{cfh}\,
    \psi_1\text{cfh}\,\psi_3;\quad \text{sfh}\,\psi_3=\text{cfh}\,\psi_1\text{cfh}\,\psi_2;\]
   \[\text{tfh}\,\psi_i\equiv\frac{\text{sfh}\,\psi_i}{\text{cfh}\,\psi_i};\quad
    \text{ctfh}\,\psi_i=1/\text{tfh}\,\psi_i.\]
As well, hold true the following obvious identities of Finsler geometry, which are analogues to the corresponding Euclidean ones:
   \[\text{sfh}\,\psi_i\text{cfh}\,\psi_i=1;\quad \text{сfh}^2\,\psi_i\text{tfh}\,\psi_i=1;\quad
   \text{sfh}^2\,\psi_i\text{ctfh}\,\psi_i=1\]
(where we have no summation for $i$), and the following identities, which have no analogues in Euclidean geometry:
   \[\text{sfh}\,\psi_1\text{sfh}\,\psi_2\text{sfh}\,\psi_3=1;\quad
    \text{tfh}\,\psi_1\text{tfh}\,\psi_2\text{tfh}\,\psi_3=1;\quad
    \text{сtfh}\,\psi_1\text{сtfh}\,\psi_2\text{сtfh}\,\psi_3=1.\]
We remark that the relative bingles, like the reciprocal ones, are conformally invariant, since they are expressed by differences between coordinates in $\H ^{\flat}.$
\section{The relation between the relative bingles and the exponential angles. Higher bingles}\label{expp}
To point out the relation between the exponential angles and the relative bingles we need to employ the bi-projection applied to the formula \bref{expug}. By means of \bref{dual} and \bref{cfh}, we infer the system of equalities:
   \begin{equation}\label{connect}\aru{\text{cfh}\psi_1=\frac{\chi_1^{2/3}}{(\chi_2\chi_3)^{1/3}}=
        e^{A_1^{2\flat}};\mm
   \text{cfh}\psi_2=\frac{\chi_2^{2/3}}{(\chi_1\chi_3)^{1/3}}=e^{A_2^{2\flat}};\mm
   \text{cfh}\psi_3=\frac{\chi_3^{2/3}}{(\chi_1\chi_2)^{1/3}}=e^{A_3^{2\flat}},}\end{equation}
which represent the needed relation between the relative bingles and the exponential angles. Here
   \[A^{2\flat}\equiv (A^{\flat})^{\flat}\]
is the element of the space $\H ^{2\flat}$ of second order bingles.\par
Hence, the system of relatie bingles $\{\psi_i\},$ of which only two are independent, due to the relation:
   \[\text{Tr}\,A^{2\flat}=\sum\limits_{i=1}^3\ln\text{cfh}\,\psi_i=0,\]
{\it characterizes both the orientation of bingles themselves, and of the elements of $\H ^{\flat}$, one relative to another, (angles in the space of angles).} This interpretation is confirmed by the following equivalent representation forms for poly-numbers:
   \begin{equation}\label{f1}A=|A|e^{\phi[A](\text{cfh}\,\psi_1\, e_1+\text{cfh}\,\psi_2\, e_2+
   \text{cfh}\,\psi_3\, e_3)}\end{equation}
{\it the exponential trigonometric} form, and
   \begin{equation}\label{f2}A=|A|e^{\phi[A]e^{A_1^{2\flat}e_1+A_2^{2\flat}e_2+A_3^{2\flat}e_3}}\end{equation}
{\it the double exponential} form. The correctness of these formulas follows from \bref{connect} and can be straightforward verified.
\section{The relation between bingles and metric invariants}
We write the expressions of reciprocal and relative bingles \bref{binggen} and \bref{bbingf} in terms of component relations: $\xi_i=B_i/A_i$:
   \begin{equation}\label{bing11}\phi[A,B]=[\ln(\xi_1^{2/3}/(\xi_2\xi_3)^{1/3})\ln(\xi_2^{2/3}/(\xi_1
    \xi_3)^{1/3})\ln(\xi_3^{2/3}/(\xi_1\xi_2)^{1/3})]^{1/3};\end{equation}
{\small
   \[\text{cfh}\,\psi_1=\frac{\ln(\xi_1^{2/3}/(\xi_2\xi_3)^{1/3})}{\phi[A,B]};\
   \text{cfh}\,\psi_2=\frac{\ln(\xi_2^{2/3}/(\xi_1\xi_3)^{1/3})}{\phi[A,B]};\
   \text{cfh}\,\psi_3=\frac{\ln(\xi_3^{2/3}/(\xi_1\xi_2)^{1/3})}{\phi[A,B]}.\]
}
For expressing these bingles in terms of conformally-invariant metric invariants:
   \[I_1\equiv\frac{1}{2}\frac{(A,A,B)}{|A|^2|B|};\quad I_2\equiv\frac{1}{2}\frac{(A,B,B)}{|A||B|^2};\quad
   I_3\equiv\frac{|A|}{|B|}=\frac{(A,A,A)}{(B,B,B)}\]
we write as well the last ones in terms of non-dimensional variables $\xi_i$:
   \[I_1=\frac{\xi_1\xi_2+\xi_2\xi_3+\xi_1\xi_3}{(\xi_1\xi_2\xi_3)^{2/3}};\quad
   I_2=\frac{\xi_1+\xi_2+\xi_3}{(\xi_1\xi_2\xi_3)^{1/3}};\quad I_3=\xi_1\xi_2\xi_3.\]
The last relations can be written in equivalent form:
   \begin{equation}\label{relatd}\xi_1+\xi_2+\xi_3=I_2I_3^{1/3};\quad
   \xi_1\xi_2+\xi_2\xi_3+\xi_1\xi_3=I_1I_3^{2/3};\quad \xi_1\xi_2\xi_3=I_3.\end{equation}
Solving the system of equations \bref{relatd} in terms of $\xi_1,\xi_2,\xi_3$ and replacing the solution in \bref{bing11}, we obtain the needed expressions of bingles in terms of metrical invariants. The system of equations \bref{relatd} has a simple algebraic interpretation. Due to the well known generalization of Viete's theorem to cubic equations we can assert that the solutions of the system \bref{relatd} are the three roots of the cubic equation:
   \[\xi^3-I_2I_3^{1/3}\xi^2+I_1I_3^{2/3}\xi-I_3=0.\]
\section{Tringles}
\subsection{Volumes in the quadratic geometries}
We remind the observations, based on which are defined volumes in the quadratic non-Euclidean geometries. At the core, lays the notion of {\it relative scalar.} We examine, as example, the standard volume form in the Euclidean space endowed with a Cartesian system of coordinates:
   \begin{equation}\label{vol0}\text{vol}_0\equiv dx_1\wedge\dots\wedge dx_n.\end{equation}
Why this form is not suitable for computing the volume in any (e.g., slant-angle or even curvilinear) system of coordinates? The answer is that for a general change of coordinates $x'=f(x)$, the form \bref{vol0} changing according to the rule:
   \begin{equation}\label{trvol1}\text{vol}'_0=\Delta_f\text{vol}_0,\end{equation}
where $\Delta_f$ is the Jacobian of the transformation $x'=f(x)$ (i.e., the determinant of the matrix $J$, containing the partial derivatives of the new coordinates relative to the old ones). From a formal algebraic point of view, the transformation law \bref{trvol1} means, that the object $\text{vol}_0$ is a relative scalar having the weight $+1$ (this multiplies with the Jacobian of the transformation at power 1). From geometric point of view, such a transformation rule means that the definition \bref{vol0} is not suitable as general definition of the volume form, since such a form has to be a scalar of weight zero (a change of sign is ignored, connected to the change of coordinates which might change the orientation of the original system of coordinates). For finding the general definition of volume, we remark that the transformation law of the metric $g$ under the before mentioned transformations has the matrix form:
   \[g'=(J^{-1})^{\text{T}}gJ^{-1},\]
where $J$ is the Jacobian matrix of the transformation, whence for determinants we infer:
   \[\det g'=\frac{\det g}{\Delta_f^2}.\]
The last formula means that the determinant of the metric is a relative scalar of weight $-2,$ and the quantity $\sqrt{\det g}$ is a relative scalar of weight $-1.$ By multiplying two relative scalars whose weight are opposite, one obtains the needed scalar of weight zero:
   \begin{equation}\label{vol1}\text{vol}\equiv \sqrt{\det g}\,dx_1\wedge\dots\wedge dx_n.\end{equation}
The formula \bref{vol1} defines the invariant volume form in any quadratic geometry in any admissible coordinate system.
\subsection{Area and volume forms in $\H$}
In non-quadratic spaces, the question regarding the shape of volume forms needs some explanation. From the usual general considerations which lead to the notion of volume, the volume form in non-quadratic spaces of dimesion $n$ should look like:
   \begin{equation}\label{vol2}\text{vol}=\mathfrak{v}\, dx_1\wedge\dots\wedge dx_n,\end{equation}
where $\mathfrak{v}$ is a relative scalar of weight $-1,$ built from the Finsler metric. For its explicit construction, we apply the general theory of invariants and covariants of poly-linear forms, which in its algebraic part is based on the theory of multi-dimensional matrices \cite{sokol}. In this work we do not need to present the general approach. We shall limit ourselves to the case of a symmetric cubic form whose components are $(G_{\alpha\beta\gamma})$ in the 2-dimensional space, which can be represented by a pair of square matrices:
   \[G=(H_1,H_2),\quad H_1\equiv G_{1\alpha\beta};\quad H_2\equiv G_{2\alpha\beta}.\]
It can be shown (\cite{sokol}), that the quantity:
   \begin{equation}\label{invar}\Delta\equiv\det\left(\begin{array}{cc}\det(H_1,H_1)& \det(H_1,H_2)\\
    \det(H_2,H_1)&\det(H_2,H_2)\end{array}\right)\end{equation}
is a relative scalar, associated to the form $G,$ and having the weight $-6.$ Here the determinant of the cubic matrix in the space of dimension 2 is given by the formula:
   \[\det G=\det(H_1,H_2)=G_{111}G_{222}-G_{112}G_{221}+G_{122}G_{211}-G_{121}G_{212}\]
In articular, when $G_{111}=G_{222}=0,$ the formula \bref{invar} leads to the expression (we ignore a non-essential constant multiplier):
   \begin{equation}\label{Delta}\Delta=G_{112}^2G_{221}^2.\end{equation}
The needed relative scalar $\mathfrak{v}$ of weight $-1$, is in this case equal to
   \begin{equation}\label{v}\mathfrak{v}=\Delta^{1/6}=(G_{112}G_{221})^{1/3}.\end{equation}
Analogously, but in a more laborious way, one can build the relative scalars of weight $-1$ for quadratic forms of higher dimensions, as well.\par
\subsection{The area form on the indicatrix and the definition of tringle}
For the metric \bref{metrS}, it is easy to build on the indicatrix, by means of formula \bref{v}, the invariant area form (the form of 2-dimensional volume):
   \begin{equation}\label{area}\text{area}_{\bm }=\frac{dX_1\wedge dX_2}{X_1X_2}.\end{equation}
We define the tringle $\Sigma(A,B,C),$ built on the triple of vectors $A,B,C$ as the area of the corresponding geodesic triangle $\Delta abc$ on the indicatrix, i.e., as the integral:
   \begin{equation}\label{tringl}\Sigma(A,B,C)=\int\limits_{\Delta abc} \frac{dX_1\wedge dX_2}{X_1X_2}.\end{equation}
It is obvious, that this tringle is the analogue of the Euclidean solid angle.
\subsection{The explicit formula of the tringle}
We displace the triangle $\Delta abc$ on the indicatrix in such a manner, that the point $a$ has the coordinates ${1,1,1}$. The coordinates of the vectors $b$ and $c$ become equal:
   \[b=\{b_1/a_1,b_2/a_2,a_1a_2/(b_1,b_2)\};\quad c=\{c_1/a_1,c_2/a_2,a_1a_2/(c_1,c_2)\}.\]
The geodesics $\Gamma_{ab}$ and $\Gamma_{ac},$ whic join $a$ with $b$ and $a$ with $c$ respectively, are parametrized as follows:
   \begin{equation}\label{geodess}\Gamma_{ab}:\ \{X_1=e^{qs},\ X_2=e^{\bar q s}\};\quad \Gamma_{ac}:\ \{Y_1=e^{q's},\
    Y_2=e^{\bar q's}\},\end{equation}
where according to \bref{formul1} we have:
   \[q=\text{cfh}\psi_1[A,B];\quad \bar q=\text{cfh}\psi_2[A,B];\quad q'=\text{cfh}\psi_1[A,C];\quad \bar
    q'=\text{cfh}\psi_2[A,C].\]
The family of parameters $q$ and $s$ can be regarded as a system of coordinates in the set of points of the triangle $\Delta abc.$ In this case, as shown before, the coordinate $q$ varies in the interval from $\text{cfh}\psi_1[A,B]$ to $\text{cfh}\psi_1[A,T],$ and the domain of variation for $s$ lays within the interval from zero to $s(q),$ where the function $s(q)$ will be specified.\par
For defining the function $s(q)$ we write the equation of the geodesic $\Gamma_{bc}$ in parametric form:
   \[Z_1=\frac{b_1}{a_1}e^{p\tau};\quad Z_2=\frac{b_2}{a_2}e^{\bar p\tau},\]
where
   \[p=\text{cfh}\psi_1[B,C];\quad \bar p=\text{cfh}\psi_2[B,C].\]
Further, we set the system of equations for finding the point of intersection of the geodesics $\Gamma_{am}$ and $\Gamma_{bc},$ where $m$ is some point on $\Gamma_{bc}$:
   \[e^{qs}=\frac{b_1}{a_1}e^{p\tau};\quad e^{\bar qs}=\frac{b_2}{a_2}e^{\bar p\tau}.\]
After minor computations, related to removal of $\tau,$ lead to the solution:
   \begin{equation}\label{solu}s(q)=\phi[A,B]\frac{\text{cfh}\,\psi_2[B,C]\text{cfh}\,\psi_1[A,B]-\text{cfh}\,
    \psi_1[B,C]\text{cfh}\,\psi_2[A,B]}{\text{cfh}\,\psi_2[B,C]q-\text{cfh}\,\psi_1[B,C]\bar q}.\end{equation}
Now for the tringle $\Sigma(A,B,C)$ we have the following chain of equations:
   \[\ba{ll}\Sigma(A,B,C)&=\dint\limits_{\Delta abc} \frac{dX_1\wedge dX_2}{X_1X_2}=\dint\limits_{\Delta abc}
    d\ln X_1\wedge d\ln X_2=\mm
    &=\dint\limits_{\Delta abc} d(qs)\wedge d(\bar qs)\dint\limits_{\Delta abc}
    \left(s\bar q-sq\frac{d\bar q}{dq}\right)\,dq\wedge ds.\ea\]
We have used the representation \bref{geodess} and the standard properties of the exterior product $\wedge.$ Replacing the derivative $d\bar q/dq$ from \bref{diffseq}, after small changes and little calculation in the integral (taking into account the relation $q\bar q(q+\bar q)=1$), we infer:
   \[\Sigma(A,B,C)=3\int\limits_{\Delta abc}\frac{s}{q(q+2\bar q)}\, dq\wedge
    ds=\int\limits_{\text{cfh}\psi_1[A,B]}^{\text{cfh}\psi_1[A,C]}\frac{dq}{q(q+2\bar
    q)}\int\limits_0^{s(q)}s\, ds.\]
Integrating by $s$, considering the formulas \bref{solu} and the properties of the function $\text{cfh}$, we get the following final formula for tringles:
   \begin{equation}\label{tringle}\Sigma(A,B,C)=\frac{3}{2}\phi^2[A,B](\text{cfh}\,\psi_1[B,C]\text{cfh}\,
    \psi_1[A,B]-\text{cfh}\,\psi_2[B,C]\text{cfh}\,\psi_2[A,B])^2\times\end{equation}
   \[\int\limits_{\text{cfh}\psi_1[A,B]}^{\text{cfh}\psi_1[A,C]}\frac{dx}{\sqrt{x^4+4x}\left(x/
    \text{cfh}\,\psi_1[B,C]-(\sqrt{x^4+4x}-x^2)/(2x\text{cfh}\,\psi_2[B,C])\right)^2}.\]
We note, that the expression \bref{tringle} is symmetric relative to cyclic permutations of the vectors $A,B,C,$, though this symmetry has been hidden because of the chosen by us local system of coordinates with origin at the point $A.$\par
The formula \bref{tringle} defines a conformally-invariant tringle, which satisfies according to its definition, the condition of additivity in the following sense. Besides the vectors $A,B,C,$ we consider the fourth vector $D,$ which satisfies one of the following properties:
   \[\ba{l}(A^{\flat}-C^{\flat})\wedge(C^{\flat}-D^{\flat})=0\quad\text{или}\mm
    (A^{\flat}-B^{\flat})\wedge(B^{\flat}-D^{\flat})=0.\ea\]
These properties mean that the points $A,C,D$ or $A,B,D$ lay inside the same plane of revolution, and the reduced to the unit sphere corresponding points $a,c,d$ or $a,b,d$ lay on one of the extremals of $\bm$. For any such point $D$ the following equality takes place:
   \begin{equation}\label{addit3}\ba{l}\Sigma(A,B,C)+\Sigma(B,C,D)=\Sigma(A,B,D)\text{или}\mm
    \Sigma(A,B,C)+\Sigma(B,C,D)=\Sigma(A,C,D),\ea\end{equation}
accordingly. This equality, in its essence, exhibits the additivity of areas on the unit sphere in $\bm$.

\end{document}